\documentclass[11pt]{amsart}
\pdfoutput=1
\usepackage[english]{babel}
\usepackage{graphicx,amsmath}
\usepackage{enumitem}
\usepackage{subfig}
\usepackage{float}
\usepackage{floatflt,young,youngtab}
\usepackage{tikz}
\theoremstyle{plain}
\numberwithin{equation}{section}
 
\newtheorem{theorem}{Theorem}[section]
\newtheorem{remark}{Remark}[section]

\newtheorem{corollary}{Corollary}[section]

\def \C {{\mathbb C}}

\def \Z {{\mathbb Z}}

\def \DKP {{\mathcal D}_{\textup{\scriptsize KP},\Gamma}}

\title[Soliton lattices of KP-II and desingularization of spectral curves in $Gr^{\mbox{\tiny TP}}(2,4)$]{Real soliton lattices of KP-II and desingularization of spectral curves: the $Gr^{\mbox{\tiny TP}}(2,4)$ case.}
\author{Simonetta Abenda}
\address{Dipartimento di Matematica, Universit\`a di Bologna, P.zza di Porta San Donato 5, I-40126 Bologna BO, ITALY
}
\email{simonetta.abenda@unibo.it
}
\author{Petr G. Grinevich}
\address{L.D.Landau Institute for Theoretical Physics,
pr. Ak Semenova 1a, Chernogolovka, 142432, Russia,
{\footnotesize pgg@landau.ac.ru}\\
Lomonosov Moscow State University,
Faculty of Mechanics and Mathematics, 
Russia, 119991, Moscow, GSP-1, 1 Leninskiye Gory, Main Building.}

\thanks{
This research has been partially supported by GNFM-INDAM and RFO University of Bologna. The second author (P.G.) partially supported by FASO Russia program No 0033-2018-0009, and by by the Russian Foundation for Basic Research, grant 17-01-00366}
\begin{document}

\begin{abstract}

We apply the general construction developed in \cite{AG2,AG3} to the first non-trivial case of $Gr^{\mbox{\tiny TP}}(2,4)$. In particular, we construct finite-gap KP-II real quasiperiodic solutions in the form of soliton lattice corresponding to a smooth genus 4 ${\mathtt M}$-curve, which is a desingularization of a reducible rational ${\mathtt M}$-curve for soliton data in $Gr^{\mbox{\tiny TP}}(2,4)$. 
\end{abstract}
\maketitle

{

\flushright \textsl{We dedicate this article to professor S. P. Novikov\\ on the occasion of his 80th birthday.}

}

\section{Introduction}
The Kadomtsev-Petviashvili (KP--II) equation \cite{KP} is the first non--trivial flow of the KP integrable hierarchy \cite{ZS} and its real solutions model, in particular, shallow water waves when the surface tension is negligible \cite{Os}:
\begin{equation}\label{eq:KP}
({-4u_t+6uu_x+u_{xxx}})_x+3u_{yy}=0,
\end{equation}
A relevant class of KP-II solutions are the complex finite--gap solutions parametrized by non--special degree $g$ divisors on non--singular algebraic curves of genus $g$ \cite{Kr1,Kr2}. Usually the analytic structure of such solutions is rather complicated.

Real regular finite-gap solutions of the  KP II equation correspond to the so-called ${\mathtt M}$-curves (see, for instance, \cite{Har,Gud,Nat,Vi}) with natural constraints on divisor positions \cite{DN}: by definition, this curve has $g+1$ real ovals, one of them contains the marked point, and each other oval contains exactly one divisor point. 

Several methods were proposed to calculate numerically real finite gap KP-II solutions, see, for example \cite{BB,DS,DFS,FK,KK1}. This activity is naturally connected to the  development of computational approaches to Riemann theta functions and their applications to integrable systems \cite{BK,DvH,DHBvHS,FK,SD,KK2}.

Recently, we \cite{AG1,AG2,AG3} proved that any multi--line real regular KP-II soliton solution may be obtained as limit of some
sequence of real quasi--periodic KP--II solutions. We stress that the relevance of obtaining \textbf{real regular} soliton solutions as limits of \textbf{real regular} finite-gap solutions was pointed out by S.P. Novikov.

Multi-line KP-II solutions are known to be represented by points in the totally non-negative part of real Grassmannians, $Gr^{\mbox{\tiny {TNN}}} (k,n)$. Our approach to assign real algebraic geomtric data to such class is constructive: we use total positivity \cite{Pin} and Postnikov parametrization \cite{Pos} of $Gr^{\mbox{\tiny {TNN}}} (k,n)$ both to associate universal reducible rational ${\mathtt M}$-curves to each positroid cell in $Gr^{\mbox{\tiny {TNN}}} (k,n)$ and, for any given such $\mathtt M$--curve, to compute the KP divisor for any soliton data in the cell.

Our results provide a way for constructing smooth ${\mathtt M}$-curves, which are are small perturbations of the rational ones, and are in agreement with the approach of \cite{Kr3} of constructing Baker--Akhiezer functions on reducible curves for degenerated finite--gap solutions. For different applications of degenerate curves to soliton theory we refer to \cite{Taim}.

We remark that the property of total positivity naturally arise in many applications, usually in connection with some reality properties of the system under study (see \cite{GK2,Pin,Kar,Lus1,Lus2,FZ2}). Recent applications to quantum field theory, regular KP soliton solutions, Josephson effect models are discussed in \cite{AGP2,CK,KW2,BG}. In particular, the relevance of total positivity in the characterization of the asymptotic behavior of these multi-line KP-II solutions in the $(x,y)$--plane for any fixed time $t$ was established in \cite{BPPP,CK}, while in \cite{KW2} the tropical limit of such solutions for $t\to\pm \infty$ has been related to the combinatorial classification of $Gr^{\mbox{\tiny TNN}} (k,n)$ of \cite{Pos} and to the cluster algebras of \cite{FZ2}.

In this paper, we apply our construction in \cite{AG3} to soliton data in the main cell $Gr^{\mbox{\tiny TP}}(2,4)$ and we explicitly construct the reducible rational curve $\Gamma({\mathcal N}_T)$ and its real and regular divisor. Then we desingularize the curve to a smooth $\mathtt M$--curve of genus 4, describe the basis of cycles and the basis of differentials, and numerically check the consistency of the construction. 

In our text we focus on real Grassamannians, but there are also natural interesting problems connected with applications of complex Grassmannians to KP theory. For recent results in this directions see \cite{BT,BT2}, \cite{KW1} and the references therein. 

\section{Finite-gap KP--II solutions}

Let us recall how to construct quasi-periodic KP--II solutions using the finite-gap approach, which was first introduced for the Korteweg-de Vries equation by S.P. Novikov \cite{Nov}. Additional details on the finite-gap approach can be found, for example, in \cite{BBEIM,DKN,Dub,Os}. The KP--II equation is part of the integrable KP hierarchy, but we consider the standard KP dynamics only, therefore we use the notation $\vec t =(t_1,t_2,t_3)=(x,y,t)$ throughout the paper. 

The finite-gap solutions of KP equation were first constructed by I.M. Krichever \cite{Kr1,Kr2}. Let $\Gamma$ be a smooth algebraic curve of genus $g$ with a marked point $P_0$ and let $\zeta^{-1}$ be a local parameter in $\Gamma$ in a neighborhood of $P_0$ such that $\zeta^{-1} (P_0)=0$. The triple $(\Gamma, P_0,\zeta^{-1})$ defines a family of exact solutions to (\ref{eq:KP}) parametrized by degree $g$ non-special divisors $\mathcal D$ defined on $\Gamma \backslash \{ P_0 \}$. 

We assume that we have a fixed canonical homological basis in $\Gamma$: $a_1$, \ldots, $a_g$, $b_1$, \ldots, $b_g$, $a_j\circ a_k=b_j\circ b_k=0$, $a_j\circ b_k=\delta_{jk}$. We denote the corresponding basis of holomorphic differentials by $\omega_1$,\ldots, $\omega_g$:
\[
\int_{a_k}\omega_l = \delta_{kl}, \quad \quad  B_{kl}=\int_{b_k}\omega_l.
\]
The symmetric matrix $B_{jk}$ with positive defined imaginary part is called the Riemann matrix for $\Gamma$.

For generic data the Baker-Akhiezer function is uniquely defined by it's analytic properties: $\tilde \Psi (P, \vec t)$ meromorphic on $\Gamma\backslash \{ P_0\}$, with simple poles at the points of the divisor ${\mathcal D}_{\textup{\scriptsize KP},\Gamma}$ and essential singularity at $P_0$ of the form
\[
{\tilde \Psi} (\zeta, \vec t) = e^{ \zeta x +\zeta^2 y +\zeta^3 t} \left( 1 - {\mathfrak w}_1({\vec t})\zeta^{-1} - \cdots
-{\mathfrak w}_l({\vec t})\zeta^{-l}  - \cdots\right). 
\]
It can be written explicitly using Its formula \cite{IM}
\begin{equation}
\label{eq:its}
\Psi(P,\vec t)=\exp\left(x\int\limits^{P} \tilde\omega_1+ y\int\limits^{P} \tilde\omega_3+ t \int\limits^{P} \tilde\omega_3
\right)\, \frac{\theta(\vec A(P) + \vec W_1 x + \vec W_2 y + \vec W_3 t + \vec C)}{\theta( \vec W_1 x + \vec W_2 y + \vec W_3 t  + \vec C)}\,
\frac{\theta(\vec C)} {\theta(\vec A(P) + \vec C)},
\end{equation}
where $\tilde\omega_j$ are the second kind meromorphic differentials with exactly one pole at $P_0$:
\[
\tilde\omega_j = \left(j \zeta^{j-1} +O\left(\frac{1}{\zeta^2}\right)\right) d\zeta,
\]
and zero $a$-cycles. $\vec W_j$ denotes the normalized vector of $b$-periods of $\tilde\omega_j$:
\[
\left(\vec W_j\right)_k=\frac{1}{2\pi i}\oint_{b_k}\tilde\omega_j,
\]
and $\vec C = -\vec A(\mathcal D)-\vec K$, where $\vec K$ is the vector of Riemann constants. We assume that integration constants in (\ref{eq:its}) are fixed by:
\[
\int\limits^{P}\tilde\omega_j = \zeta^{j} + o(1).
\]
In our text we use the same normalization of the Riemann theta-functions as in the book \cite{Mum}:
\[
\theta(z) = \theta(\vec z| B) = \sum\limits_{{n_j \in \Z} \atop {j=1,\ldots,g}} \exp\left[\pi i \sum_{k,l=1}^g B_{kl}n_kn_l + 2\pi i \sum_{k=1}^g z_k n_k \right].
\]

Let us denote the expansion coefficients of $\tilde\omega_j$ at $P_0$ by $\hat\omega_{jk}$:
\[
\tilde\omega_j = \left(j \zeta^{j-1} +\sum\limits_{k=1}^{\infty}\frac{\hat\omega_{jk}}{\zeta^{k+1}} \right) d\zeta.
\]
From Riemann bilinear relations (see, for example, \cite{Spr}) it follows that the matrix $\hat\omega_{jk}$ is symmetric: $\hat\omega_{jk}=\hat\omega_{kj}$. From  Riemann bilinear relations it also
follows that near the point $P_0$ we have:
\[
\vec A(P) = -\frac{\vec W_1}{\zeta}-\frac{\vec W_2}{2\zeta^2}- \frac{\vec W_3}{3\zeta^3}+O\left(\frac{1}{\zeta^4} \right), \ \ \zeta=\zeta(P).
\]

The Baker-Akhiezer function $\Psi(P,\vec t)$ is a common eigenfunction for a pair of operators
\[
(\partial_y -B_2)\Psi(P,\vec t) = (\partial_t -B_3)\Psi(P,\vec t) =0,
\]
where
\begin{equation}
\label{eq:zerocurv}
B_2 = \partial_x^2 + u(\vec t), \ \ B_3= \partial_x^3 + \frac{3}{4}\bigg(\partial_x\circ u(\vec t)+  u(\vec t)\circ  \partial_x\bigg) + w(\vec t), \ \  \partial_xw(\vec t) = \frac{3}{4}  \partial_yu(\vec t).  
\end{equation}
Operators $(\partial_y -B_2)$, $(\partial_t -B_3)$ form the zero-curvature representation for KP-II \cite{Druma,ZS}, therefore the function $u(x,y,t)$ solves the KP-II equation (\ref{eq:KP}). Its explicit representation is given by the Its-Matveev formula:
\begin{equation}\label{eq:fingap}
u(\vec t) = 2\partial_x^2 \log \theta (x \vec W_1+y \vec W_2+t \vec W_3 + \vec C)+2 \hat \omega_{11}.
\end{equation}

To construct real KP--II solutions one has to impose reality condition on the spectral data, more precisely, $\Gamma$ must be a real curve, i.e. admit an antiholomorphic involution (complex conjugation)
\[
\sigma : \Gamma \to \Gamma, \quad \sigma^2 =1,
\]
such that 
\begin{enumerate}
\item $\sigma(P_0)=P_0$;
\item $\sigma(\zeta) = \bar\zeta$;
\item $\sigma({\mathcal D}) = {\mathcal D}$. This property does not imply that all points of ${\mathcal D}$ are fixed points of $\sigma$, since the involution  $\sigma$ may interchange some of them.
\end{enumerate}
The necessary and sufficient conditions for the smoothness of these real solution (\ref{eq:fingap}) associated with smooth curve $\Gamma$ of genus $g$ have been proven in \cite{DN}: 
\begin{enumerate}
\item $\Gamma$ must be an $\mathtt M$--curve, i.e. the set of fixed points of ${\sigma}$ consists of $g+1$ ovals, $\Omega_0,\Omega_1,\dots,\Omega_g$. 
\item If we denote by $\Omega_0$ the oval containing $P_0$, then each other oval contains exactly one divisor point.
\end{enumerate}
Ovals $\Omega_j$ are called ``fixed'' or ``real''. The set of real ovals divides  $\Gamma$ into two connected components. 
Each of these components is homeomorphic to a sphere with $g+1$ holes. We call the ovals $\Omega_j$, $j>0$,  ``finite''.

\begin{figure}[H]
  \centering
  {\includegraphics[width=0.49\textwidth]{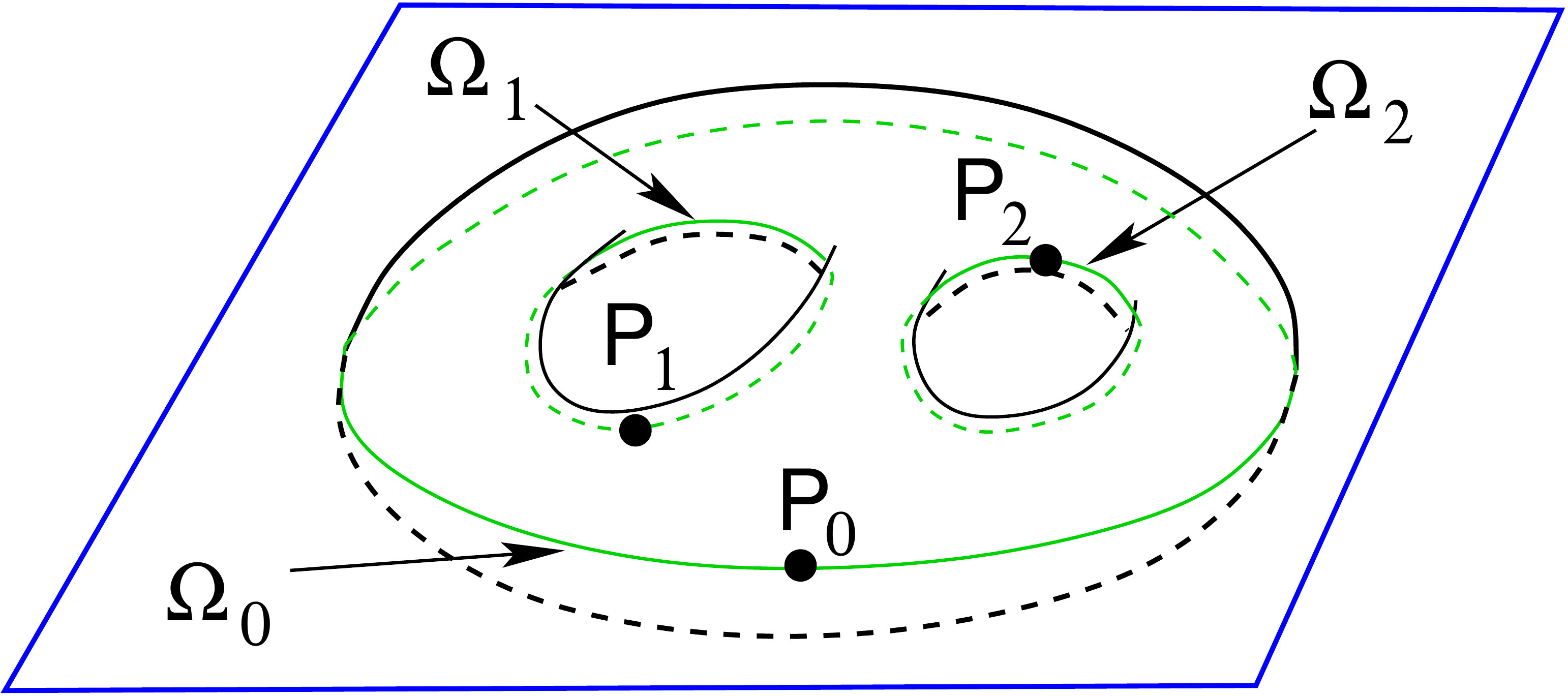}}
  {\includegraphics[width=0.49\textwidth]{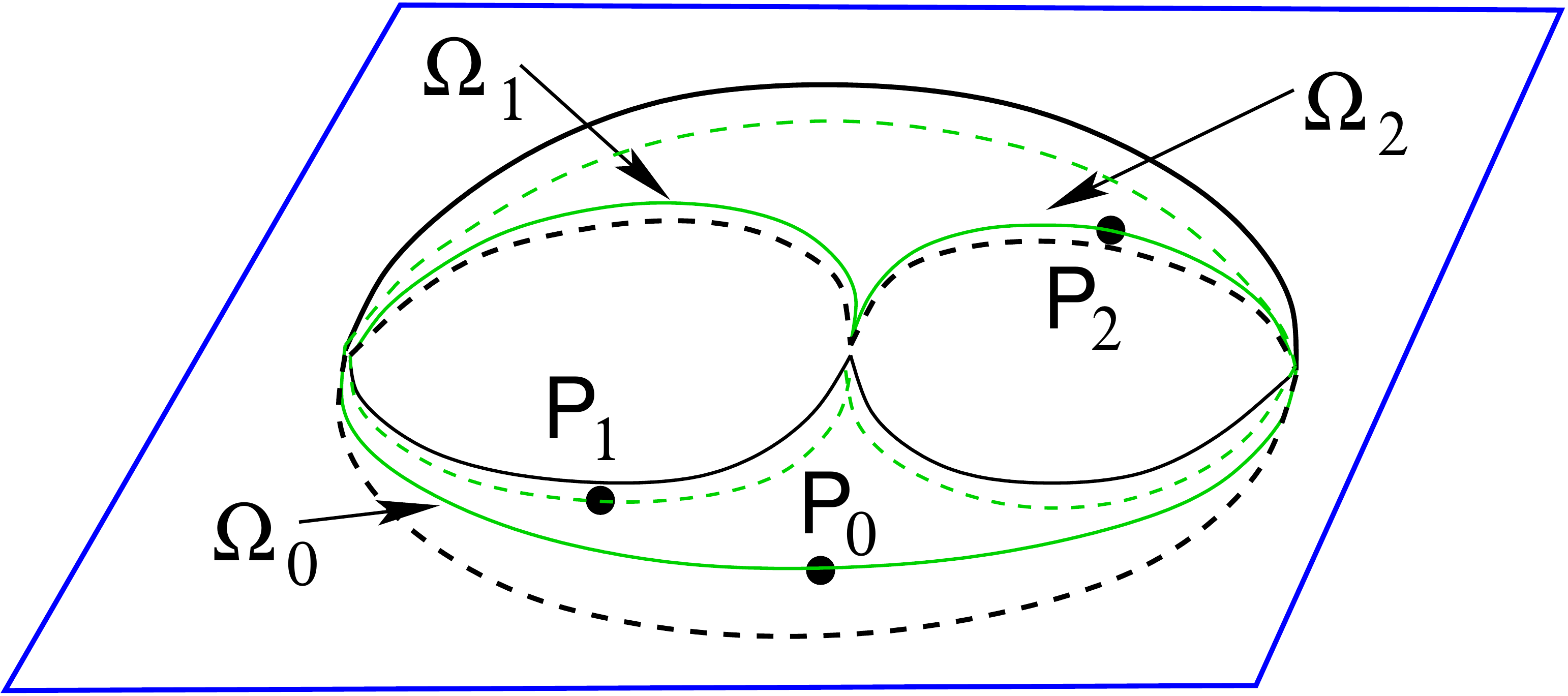}}
\caption{\small{\sl Left: A degree 2 divisor on a regular $\mathtt M$-curve of genus $g=2$ satisfying the reality and regularity conditions in \cite{DN}. Right: In the solitonic limit, the spectral curve is a reducible rational $\mathtt M$--curve and the divisor satisfying the reality and regularity conditions represents soliton data in $Gr^{\mbox{\tiny TP}} (1,3)$.}}\label{fig:regM_g2}        
\end{figure}

\begin{remark}
\label{rem:ovals}
To have a good convergence of the theta-series, it is convenient to choose the following homological basis: $b_j=\Omega_j$, and the $a$-cycles are purely imaginary $\sigma(a_j)=-a_j$, $j\in[g]$. Note, this convention is opposite to the choice in \cite{DN}.
\end{remark}
Finally, let us remark that these solutions are quasi--periodic solutions for real $x$, $y$, $t$. 

In our normalization the matrix $B$ and the vectors $\vec W_1$, $\vec W_2$, $\vec W_3$ are purely imaginary, and the expansion coefficients $\hat\omega_{jk}$ are pure real. The condition that each finite fixed oval contains exactly one divisor point means that $\vec C$ is pure imaginary (see \cite{DN}).

For instance, in Figure~\ref{fig:regM_g2}[left] the regular curve has genus 2, the involution is the orthogonal reflection w.r.t. the horizontal plane and it fixes three oval. The reality and regularity conditions from \cite{DN} impose that the essential singularity $P_0$ of the wave--function belongs to one such oval and that there is one simple pole $P_i$, $i=1,2$, in each of the remaining ovals.

\section{Real regular bounded KP--II multi--line soliton solutions}

Another interesting class of KP--II solutions, which can be interpreted as a nonlinear superposition of line solitons, can be constructed using the dressing procedure \cite{ZS,S},  or, equivalently, the Darboux transfromations \cite{Mat}. Let us recall the main formulas.

Let $f^{(i)}$, $i\in[k]$, be a collection of linearly independent solutions to the heat hierarchy $\partial_{t_l} f^{(i)} = \partial_x^l f^{(i)}$, for $l=2,3$. 

Let ${\mathfrak D}$ be the $k$-th order ordinary differential operator
\begin{equation}\label{eq:D}
{\mathfrak D}\equiv W\partial_x^k =\partial_x^k -{\mathfrak w}_1 (\vec t)\partial_x^{k-1} -\cdots - {\mathfrak w}_k(\vec t),
\end{equation}
uniquely defined imposing that
\begin{equation}
\label{eq:def_D}
{\mathfrak D} f^{(i)} =0, \quad i\in [k].
\end{equation}

The unnormalized dressed wave function
\begin{equation}
\label{eq:Satowf}
\psi^{(0)} (\zeta; \vec t) = {\mathfrak D}  e^{\theta (\zeta, {\vec t})} = \left( \zeta^k - {\mathfrak w}_1({\vec t})\zeta^{k-1} - \cdots
- {\mathfrak w}_k({\vec t}) \right) e^{\theta (\zeta, {\vec t})},
\end{equation}
(here $\theta(\zeta, \vec t) = \zeta x + \zeta^2 y +\zeta^3 t$),
is a common eigenfunction for the pair of operators forming the zero-curvature representation (\ref{eq:zerocurv})
\[
(\partial_y-B_2) \psi^{(0)} (\zeta; \vec t) = (\partial_t-B_3) \psi^{(0)} (\zeta; \vec t) =0,
\]
therefore $u(\vec t)= 2\partial_x  {\mathfrak w}_1({\vec t})$ solves the KP--II equation  (\ref{eq:KP}). The corresponding Sato $\tau$-function is
\[
{\displaystyle \tau (\vec t)} \displaystyle = \mbox{ Wr}_x (f^{(1)},\dots, f^{(k)}),
\]
and we also have \cite{Mat}
\begin{equation}\label{eq:sol}
u (\vec t) = 2 \partial_x^2 \log ({\tau (\vec t)}).
\end{equation}

Multi-line soliton solutions are obtained in the special case when 
\[
f^{(i)} (\vec t)=\sum_{j=1}^n A^i_j E_j (\vec t), \quad i\in [k], \quad\quad E_j (\vec t) = \exp(\kappa_j x + \kappa_j^2 y +\kappa_j^3 t).
\]
In this case
\[
{\displaystyle \tau (\vec t)} = \sum\limits_{1\le j_1<\cdots<j_k\le n} \Delta_{(j_1,\dots,j_k)} (A) \prod_{1\le r<s\le k} (\kappa_{j_s} -\kappa_{j_r}) \prod_{l=1}^k E_{j_l} (\vec t),
\]
where $\Delta_{(j_1,\dots,j_k)} (A)$ are the maximal minors of the $k\times n$ matrix $A=(A^i_j)$, with ordered columns $1\le j_1<\cdots < j_k\le n$. 

To obtain real regular uniformly bounded multi-line solutions it is sufficient \cite{Mal} and necessary \cite{KW1} that a) the $n$ phases $\kappa_j$ are real; b) if the phases are ordered  as $\kappa_1<\cdots < \kappa_n$, then all maximal minors of $A$ are non--negative, $\Delta_{(j_1,\dots,j_k)} (A)\ge 0$. 

Any linear recombination of the rows of $A$ generates the same $u(\vec t)$; therefore regular uniformly bounded solutions (\ref{eq:sol}) are parametrized by points $[A]$ in the totally non--negative Grassmannian $Gr^{\mbox{\tiny TNN}} (k,n)= GL_{\mathbb{R}}^+ (k) \backslash Mat_{\mathbb{R}}^{\mbox{\tiny TNN}} (k,n)$ \cite{KW1}. Here $Mat_{\mathbb{R}}^{\mbox{\tiny TNN}} (k,n)$ denotes the set of real $k\times n$ matrices with nonnegative maximal minors.  

In this paper we consider the special case in which $[A]\in Gr^{\mbox{\tiny TP}} (2,4)$. $Gr^{\mbox{\tiny TP}}(2,4)$ is the main cell in $Gr^{\mbox{\tiny TNN}} (2,4)$ and its elements $[A]$ are equivalence classes of real $2\times 4$ matrices $A$ with all maximal minors positive. Such matrices are parametrized by four positive numbers, $w_{ij}$, $i=1,2$, $j=3,4$, and may be represented in the reduced row echelon form (RREF),
\begin{equation}
\label{eq:RREF}
A = \left( \begin{array}{cccc}
1 & 0 & - w_{13} & -w_{13} (w_{14}+w_{24})\\
0 & 1 & w_{23}   & w_{23}w_{24}
\end{array}
\right).
\end{equation}
The heat hierarchy solutions are then
\[
\begin{array}{l}
f^{(1)} (\vec t) = E_1 (\vec t) -w_{13} E_3 (\vec t) -w_{13}(w_{14}+w_{24}) E_4(\vec t), \\
f^{(2)} (\vec t) = E_2 (\vec t) +w_{23} E_3 (\vec t) +w_{23}w_{24} E_4(\vec t),
\end{array}
\]
and the KP--II solution is 
\[
u(\vec t) = 2\partial_x^2 \log\left( f^{(2)}\partial_x f^{(1)}-f^{(1)}\partial_xf^{(2)}\right).
\]

\section{The construction of the reducible curve $\Gamma({\mathcal N}_T)$ and of the KP--II divisor for soliton data in $Gr^{\mbox{\tiny KP}}(k,n)$}

In this section we briefly summarize our construction in \cite{AG3} and, for simplicity, we restrict ourselves to soliton data $[A]\in Gr^{\mbox{\tiny TP}} (k,n)$.

The Darboux transformation in (\ref{eq:D}) provides the following spectral data: a copy of $\mathbb{CP}^1$, $\Gamma_0$, with a marked point $P_0$, $n$ cusps $\kappa_1,\dots,\kappa_n$, and a real $k$ point divisor $\mathcal D^{(0)} = \{ P_l \, ; l\in [k] \}\subset \Gamma_0 \backslash \{ P_0 \}$, such that the $\zeta$--coordinate $\zeta(P_l)=\gamma_l\in [\kappa_1 , \kappa_n]$ satisfy
\begin{equation}
\label{eq:Satodiv}
(\gamma_l)^k - {\mathfrak w}_1 (\vec t_0) (\gamma_l)^{k-1} - \cdots - {\mathfrak w}_{k-1} (\vec t_0) \gamma_l-{\mathfrak w}_k (\vec t_0) = 0 ,\quad\quad l\in [k],
\end{equation}
and the initial time $\vec t_0$ will be specified later.

However, if we fix $\mathcal K$ and let $[A]$ vary in $Gr^{\mbox{\tiny TP}} (k,n)$, we get a $k(n-k)$--dimensional family of real bounded multiline solitons. Then, following Krichever \cite{Kr3}, we expect that they may be locally parametrized via $k(n-k)$ point divisors on some reducible curve $\Gamma$ of which $\Gamma_0$ is a rational component.

Moreover, the sufficient part of Dubrovin and Natanzon's proof \cite{DN} also holds when the algebraic $\mathtt M$-curve is singular. Since 
the multi-soliton solutions associated to points in $Gr^{\mbox{\tiny TNN}} (k,n)$ are real bounded and regular for all $\vec t$, it is natural to expect that they may be associated 
to real algebraic-geometric data on reducible curves which are rational degenerations of regular $\mathtt M$--curves. 

The above ansatz has been proven in \cite{AG1,AG2,AG3} for all soliton data in $Gr^{\mbox{\tiny TNN}}(k,n)$. 
In our approach \cite{AG1,AG2,AG3}, $\Gamma_0$ is a component of a reducible curve $\Gamma$ and $\mathcal D^{(0)}$ is the restriction to $\Gamma_0$ of the Krichever real regular divisor on $\Gamma\backslash \{P_0\}$. 
We show the simplest example of our construction in Figure~\ref{fig:regM_g2}[right] (see \cite{A1,AG1} for necessary details): soliton data in $Gr^{\mbox{\tiny TP}} (1,3)$ are parametrized by divisors satisfying the reality and regularity conditions in \cite{DN} on a rational degeneration of a real genus 2 hyperelliptic curve.

In particular, in \cite{AG3}, we use the relations between positive Grassmannians and networks established in \cite{Pos}, to construct 
a reducible curve $\Gamma({\mathcal N}_T)$, which is a rational degeneration of a smooth $\mathtt M$--curve using the Le-network ${\mathcal N}_T$ representing $[A]$. Since the Le--network provides a minimal parametrization of the positroid cell, by construction $\Gamma({\mathcal N}_T)$ is the rational degeneration of a smooth curve of minimal genus equal to the dimension of
the positroid cell for $[A]$. We refer to \cite{Pos} for more details on totally non--negative Grassmannians and their combinatorial classification. 

$\Gamma({\mathcal N}_T)$ is obtained reflecting the graph of ${\mathcal N}_T$ with respect to a line orthogonal to the one containing the boundary vertexes via the following natural correspondence: internal vertexes, edges and faces  of ${\mathcal N}_T$ respectively correspond to $\mathbb{CP}^1$ components, double points and ovals of $\Gamma({\mathcal N}_T)$. The boundary of the disk containing ${\mathcal N}_T$ is $\Gamma_0$, while the boundary vertexes are the phases in $\mathcal K$. Let us remark that this construction is rather close to the construction connecting a rational curve to its dual graph.  We illustrate such correspondence in Table \ref{table:LeG}.

\begin{table}[H]
\caption{The graph of ${\mathcal N}_T$  vs the degenerate rational curve $\Gamma$} 
\centering
\begin{tabular}{|c|c|}
\hline\hline
Graph & Degenerate $\mathtt M$--curve\\[0.5ex]
\hline
Boundary of disk & Copy of $\mathbb{CP}^1$ denoted $\Gamma_0$ \\
Boundary vertex $b_l$             & Marked point $\kappa_l$ on  $\Gamma_0$\\
Black vertex   $V^{\prime}_{ij}$  & Copy of $\mathbb{CP}^1$ denoted $\Sigma_{ij}$\\
White vertex   $V_{ij}$           & Copy of $\mathbb{CP}^1$ denoted $\Gamma_{ij}$\\
Internal Edge                     & Double point\\
Face                              & Oval\\ [1ex]
\hline
\end{tabular}
\label{table:LeG}
\end{table}

\begin{figure}[H]
\centering{
\includegraphics[width=0.38\textwidth]{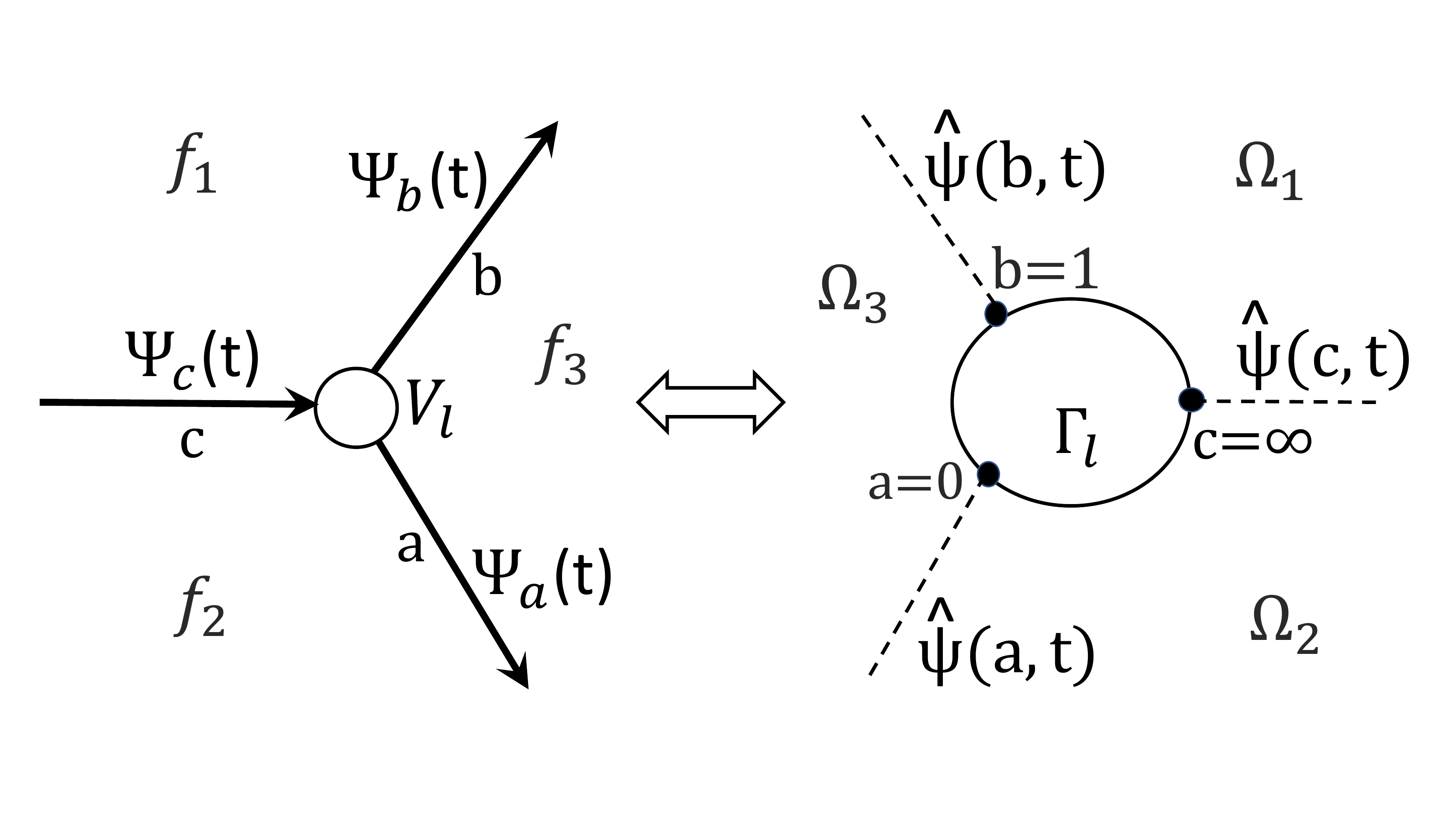}}
\caption{\small{\sl The rational component [right] is obtained reflecting the graph [left] w.r.t. a vertical line using Table \ref{table:LeG}. Each $\mathbb{RP}^1$ component is oriented clockwise and the coordinates of the marked points [right] correspond to the anticlockwise enumeration of edges at vertexes [left]. We also show the correspondence between the  edge wavefunction $\Psi_e(\vec t)$ on the edge $e$ at $V_r$ [left] and the normalized edge wavefunction $\hat \psi(e,\vec t)$ at the marked point $e\in \Gamma_r$ [right].}}
\label{fig:coor}
\end{figure}

We graphically represent $\Gamma$ by its topological model, and visualize double points of $\Gamma$ by dotted lines. We also introduce a global coordinate on each $\mathbb{CP}^1$ component. In Figure \ref{fig:coor}, we illustrate such correspondence for components represented by trivalent white vertexes, since, in our construction, the normalized wavefunction may have untrivial dependence on the spectral parameter only at these components. In Figure \ref{fig:Gr24_net} we compare the Le--network and the curve for soliton data in $Gr^{\mbox{\tiny TP}}(2,4)$.

To each edge $e\in {\mathcal N}_T$ we assign a vacuum edge wavefunction $\Phi_e (\vec t)$ and its dressing $\Psi_e (\vec t) = {\mathfrak D} \Phi_e (\vec t)$
using linear recurrence at each vertex (see \cite{AG3} for more details on the construction). Using the same linear relations, we assign a dressed divisor number to each white vertex. 

We then normalize the dressed edge wavefunction at the double points of each $\mathbb{CP}^1$ component
\begin{equation}
\label{eq:normal}
\hat\psi(e,\vec t)= \frac{\Psi_e(\vec t)}{\Psi_e(\vec t_0)},
\end{equation}
where $\vec t_0$ is chosen so that all denominators are different from zero \cite{AG3}. In the left-hand side of (\ref{eq:normal}) $e$ denotes the double point in $\Gamma$, corresponding to the edge $e$ in the Le-network.  By construction the normalized dressed edge wavefunction $\hat \psi(e,\vec t)$ takes the same value at all the marked points of each $\mathbb{CP}^1$ component corresponding either to a black vertex or to a bivalent white vertex, $V_l$; therefore on any such rational component we analytically extend $\hat \psi$ to a wavefunction constant with respect to the spectral parameter. At each trivalent white vertex $V_l$ the normalized dressed edge wavefunction $\hat \psi(P,\vec t)$ takes three distinct values at the double points for generic $\vec t$; therefore we extend it on $\Gamma_l$ to a degree one meromorphic function and prove that the dressed divisor number assigned to $V_l$ is the coordinate of the KP--II pole divisor point on $\Gamma_l$.

Finally any change of orientation of the network ${\mathcal N}_T$ induces a well--defined transformation of coordinates on the components of $\Gamma({\mathcal N}_T)$ which leaves invariant the positions of the double points, the value of the normalized dressed wavefunction at the double points and the pole divisor \cite{AG2}. The following theorem is a particular case of the more general construction presented in \cite{AG2,AG3}.

\begin{theorem}\label{theo:AG}
Let $({\mathcal K}, [A])$ be given soliton data with ${\mathcal K} = \{ \kappa_1 < \cdots <\kappa_n\}$ and $[A]\in Gr^{\mbox{\tiny TP}}(k,n)$. Let ${\mathcal N}_T$ be the Le--network representing $[A]$ and let $\Gamma({\mathcal N}_T)$ be the corresponding reducible curve constructed in \cite{AG2}. Let $\DKP$ and $\hat \psi (P,\vec t)$ respectively be the dressed divisor and the normalized dressed wavefunction on $\Gamma({\mathcal N}_T)$. Then
\begin{enumerate}
\item $\Gamma({\mathcal N}_T)$ is the rational degeneration of a smooth $\mathtt M$--curve of genus $g$ equal to the dimension $k(n-k)$ of $Gr^{\mbox{\tiny TP}}(k,n)$: $g=k(n-k)$;
\item $\DKP$ is an effective degree $g$ divisor and depends only on the soliton data $\DKP=\DKP(\mathcal K,[A])$, in particular it does not depend on $x$, $y$, $t$;
\item $k$ poles of $\DKP$ coincide with the Sato divisor and belong to $\Gamma_0$; the remaining $g-k$ poles belong to the copies of $\mathbb{CP}^1$ corresponding to the trivalent white vertexes of ${\mathcal N}_T$;
\item The marked point $P_0$ (essential singularity) belongs to one oval of $\Gamma({\mathcal N}_T)$ and any other oval of $\Gamma({\mathcal N}_T)$ contains exactly one divisor point of $\DKP$; 
\item $\hat \psi (P, \vec t)$ is meromorphic in $P$ on $\Gamma({\mathcal N}_T)\backslash\{P_0\}$ and regular in $\vec t$, and its divisor satisfies
$( \hat \psi (P, \vec t)) + \DKP \ge 0$ for all $\vec t$;
\item $\hat \psi(P, \vec t)$ satisfies the gluing conditions at all double points of $\Gamma({\mathcal N}_T)$ for all $\vec t$;
\item $\hat \psi (P, \vec t)$ is independent on $P$ on each copy of $\mathbb{CP}^1$ corresponding either to a black vertex or a bivalent white vertex in ${\mathcal N}_T$ .
\item $\hat \psi (P, \vec t)$ is meromorphic in $P$ of degree $\le 1$ on each copy of $\mathbb{CP}^1$ corresponding to a trivalent white vertex in ${\mathcal N}_T$.
\end{enumerate}
\end{theorem} 

\begin{corollary}\cite{AG2}
$\DKP$ and $\hat \psi(P,\vec t)$ respectively are the KP--II divisor and the KP--II wavefunction on $\Gamma(\mathcal N_T)$ for the soliton data 
$({\mathcal K}, [A])$.
In particular, $\DKP$ satisfies the reality and regularity conditions of \cite{DN}.
\end{corollary}

\section{The KP divisor for soliton data in $Gr^{\mbox{\tiny TP}}(2,4)$}

\begin{figure}[H]
\centering{
\includegraphics[width=0.48\textwidth]{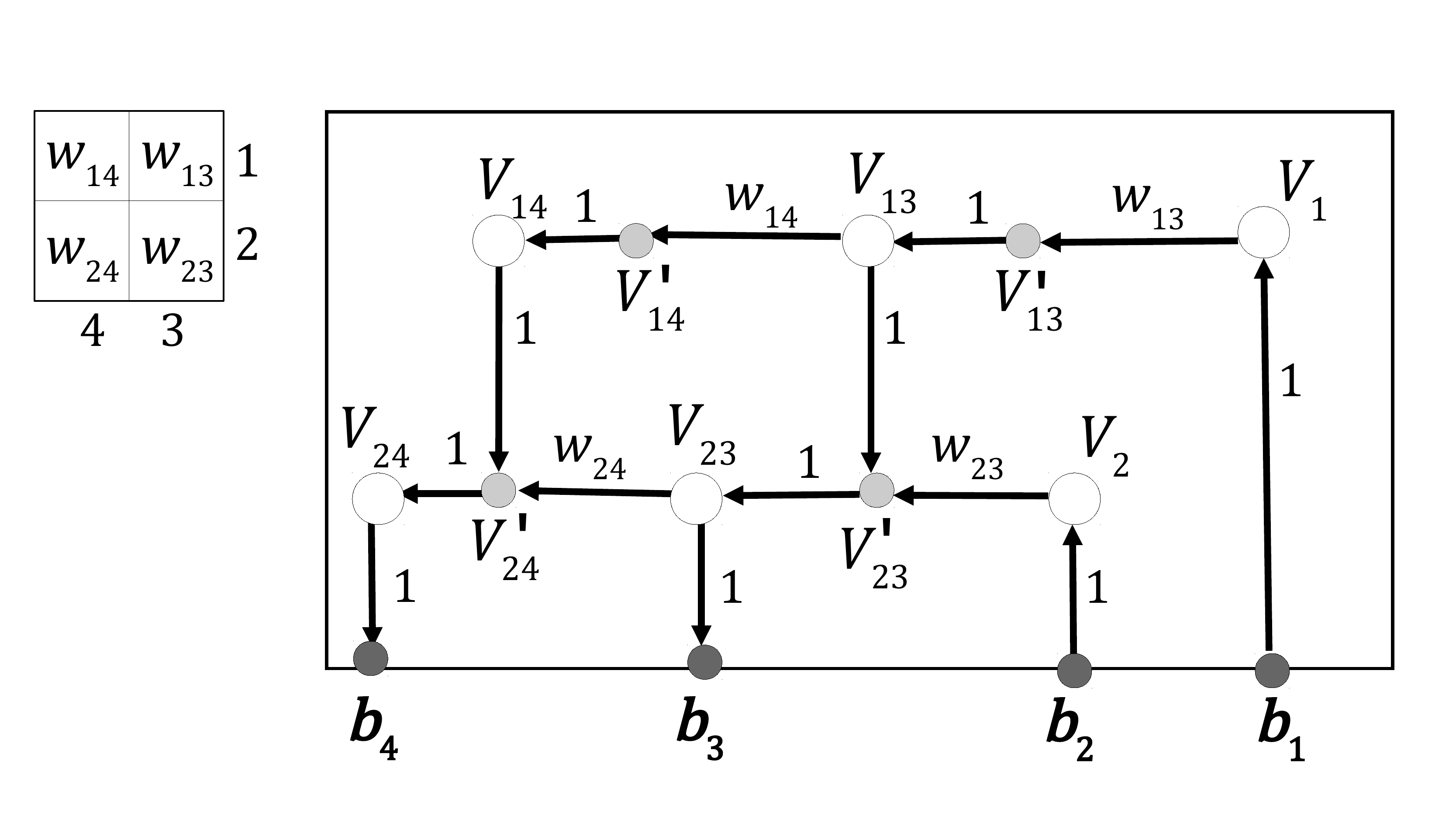}
\hfill
\includegraphics[width=0.49\textwidth]{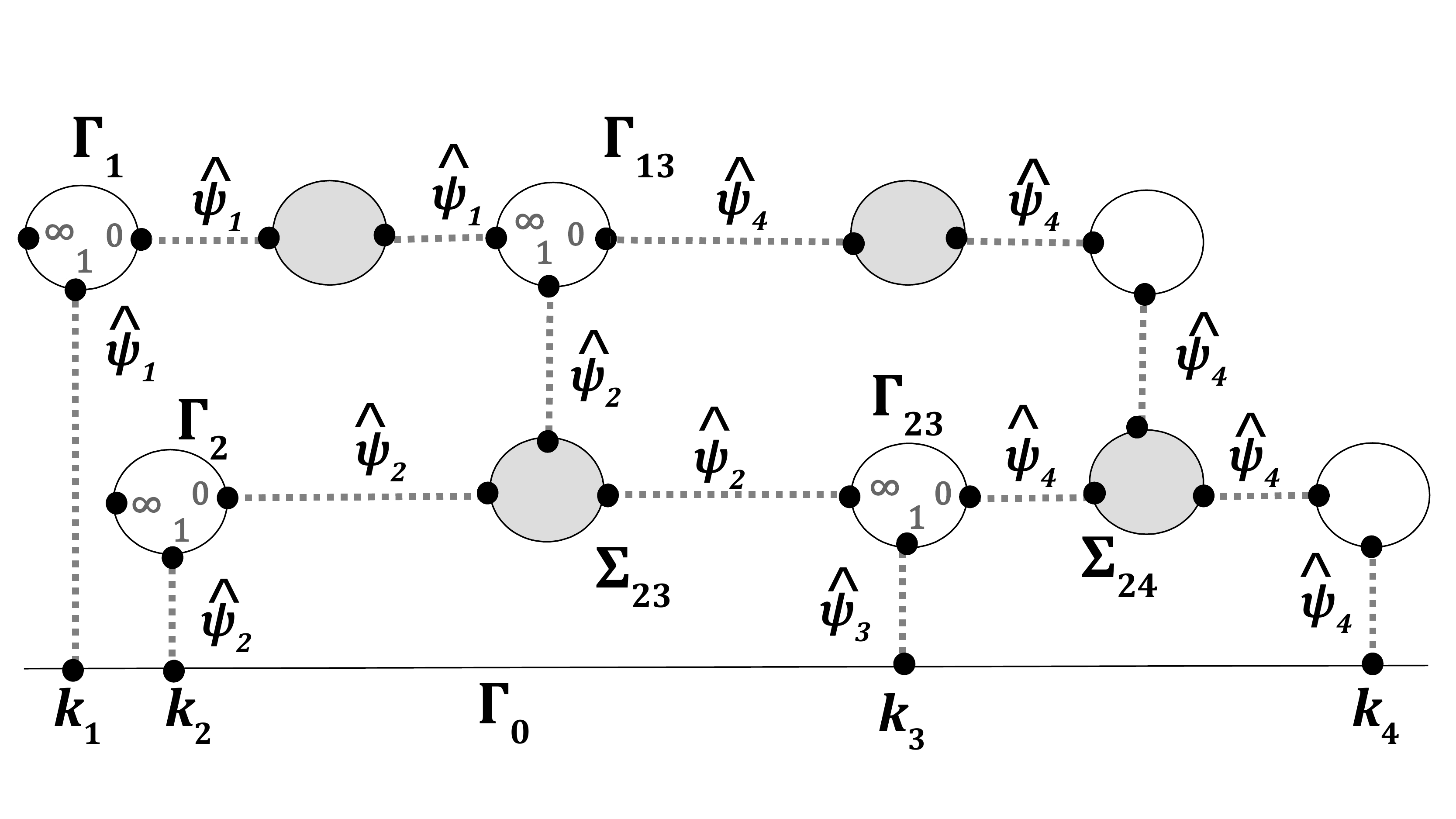}
\includegraphics[width=0.48\textwidth]{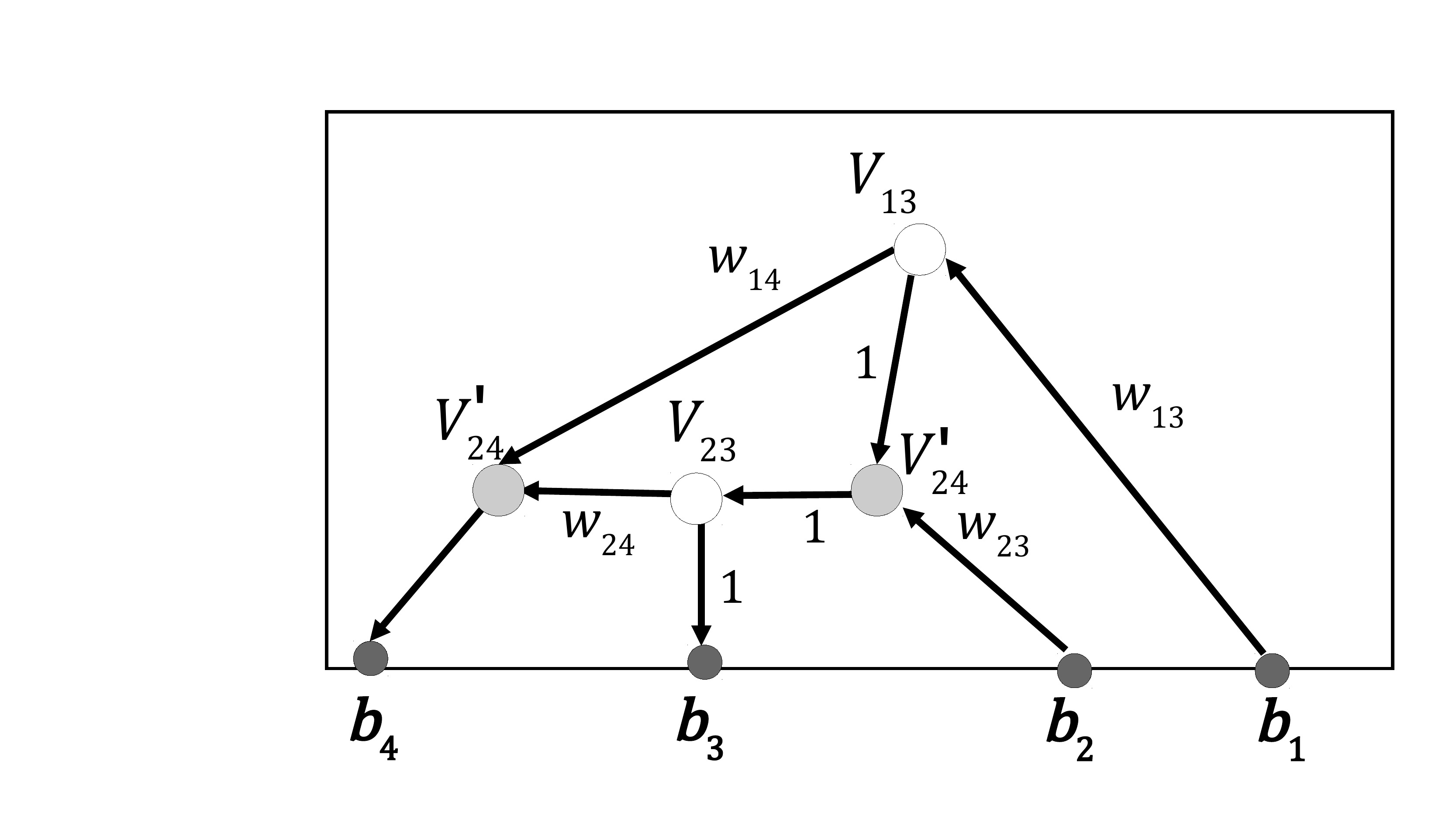}
\hfill
\includegraphics[width=0.49\textwidth]{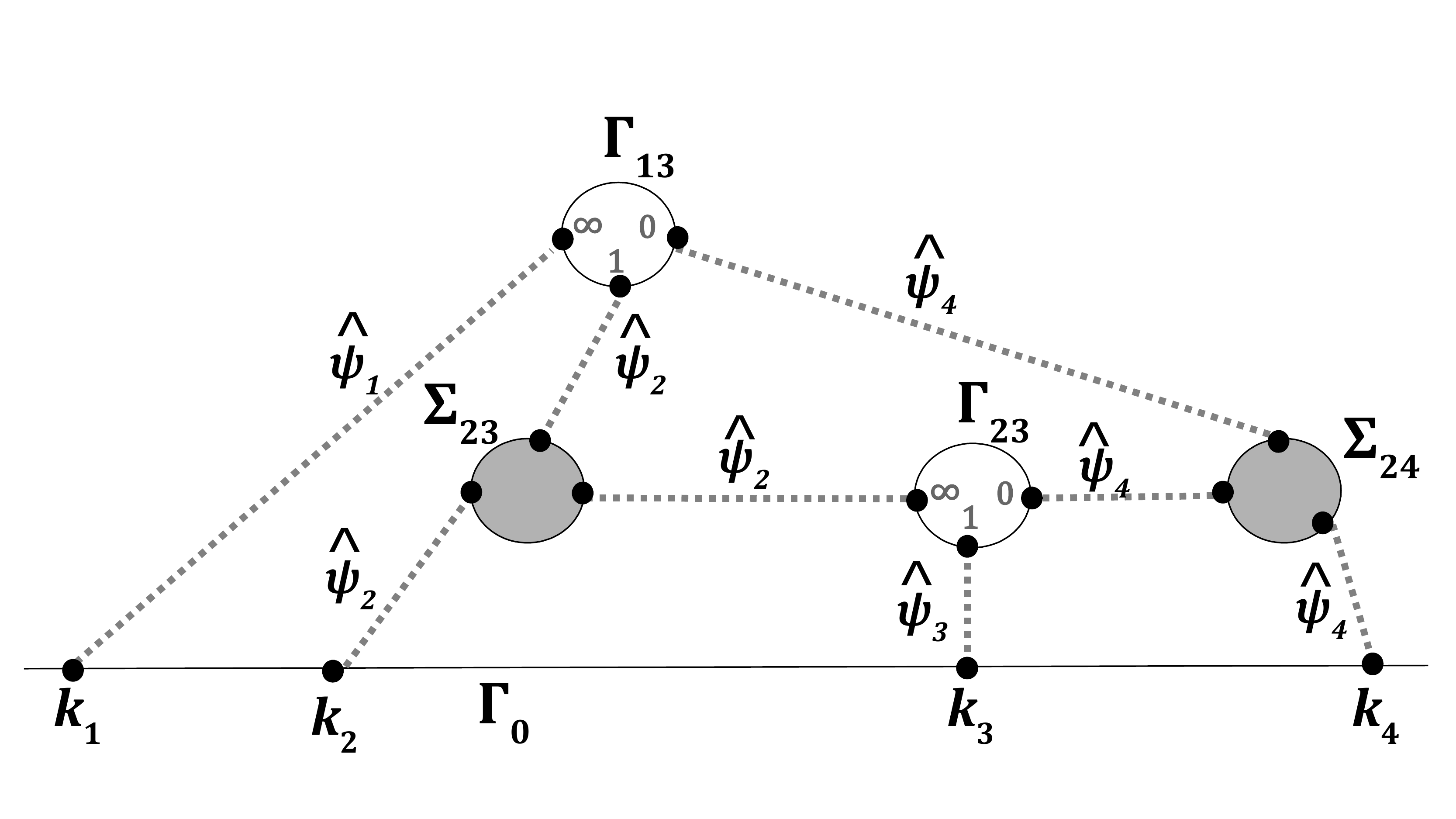}
}
\caption{\small{\sl Top: The Le-network ${\mathcal N}_T$ [left] and the topological model of the spectral curve $\Gamma ({\mathcal N}_T)$ [right] for the point $[A]\in Gr^{\mbox{\tiny TP}}(2,4)$ with $A$ as in (\ref{eq:RREF}). Bottom: the reduced Le-network ${\mathcal N}_{T,{\mbox{\scriptsize red}}}$ [left] and the topological model of the spectral curve $\Gamma ({\mathcal N}_{T,{\mbox{\scriptsize red}}})$ [right]. On both curves, double points are represented as dotted segments and $\hat \psi_l \equiv \hat\psi_l (\vec t)$ is as in (\ref{eq:hat_psi}).}}
\label{fig:Gr24_net}
\end{figure}

Let us briefly illustrate the construction of the wavefunction and of the divisor in the case of soliton data $[A]\in Gr^{\mbox{\tiny TP}}(2,4)$. We have considered the same example also in \cite{A2,AG3} with different purposes: in \cite{A2} we show the compatibility of the asymptotic behavior of its KP-II zero divisor with that of the corresponding soliton solution classified in \cite{CK}, while in \cite{AG3} we compare the construction of the KP divisor with the approach in \cite{AG1}.

In Figure \ref{fig:Gr24_net} [left], we respectively show the acyclically oriented Le--network ${\mathcal N}_T$ representing $[A]\in Gr^{\mbox{\tiny TP}}(2,4)$ and its reduction ${\mathcal N}_{T, {\mbox{\scriptsize red}}}$  obtained eliminating all bivalent vertexes. We remark that the weights on the edges are exactly the parameters in  (\ref{eq:RREF}), and, in agreement with the general construction of \cite{Pos}, the maximal minors $\Delta_{j_1,j_2}$  of $A$ in (\ref{eq:RREF}) may be computed as sums of the total weights of all non-intersecting pairs of oriented paths from the boundary sources $b_1$, $b_2$ to the boundary vertexes $b_{j_1}$, $b_{j_2}$.

The ratio of such minors is the boundary measurement map introduced in \cite{Pos}, and it is invariant with respect to the reduction ${\mathcal N}_T\rightarrow{\mathcal N}_{T, {\mbox{\scriptsize red}}}$. The counterpart of this property in our construction \cite{AG2} is the invariance of the normalized KP-II edge wavefunction $\hat \psi (P, \vec t)$ at the corresponding double points on $\Gamma({\mathcal N}_T)$ and on $\Gamma({\mathcal N}_{T,{\mbox{\scriptsize red}}})$ (see Figure \ref{fig:Gr24_net} [right]). Therefore the KP--divisor points are the same in both cases.

Using Theorem~\ref{theo:AG} (see also \cite{A2,AG3}), it is easy to verify that, in this special case, the KP wavefunction  may take only four possible values at the marked points:
\begin{equation}\label{eq:hat_psi}
\hat \psi_l (\vec t) = \frac{{\mathfrak D} e^{\theta_l(\vec t)}}{{\mathfrak D} e^{\theta_l(\vec t_0)}}, \quad \quad l\in [4],
\end{equation}
where $\theta_l(\vec t) = \kappa_l x+  \kappa_l^2 y+ \kappa_l^3 t$. In Figure~\ref{fig:Gr24_net} [right] we show which double point carries which of the above values of $\hat \psi$. We remark that the value $\hat \psi_l (\vec t)$ at each marked point is independent of the choice of local coordinates on the components, {\sl i.e.} of the orientation in the corresponding network \cite{AG2}. 
The position of the KP divisor points of $\DKP$ is the same both for $\Gamma=\Gamma ({\mathcal N}_T)$ and $\Gamma=\Gamma ({\mathcal N}_{T,{\mbox{\scriptsize red}}})$ and the local coordinate of each divisor point may be computed using the correspondence with the orientation of the network established in \cite{AG2,AG3}. 

\begin{figure}[H]
\centering{
\includegraphics[width=0.3\textwidth]{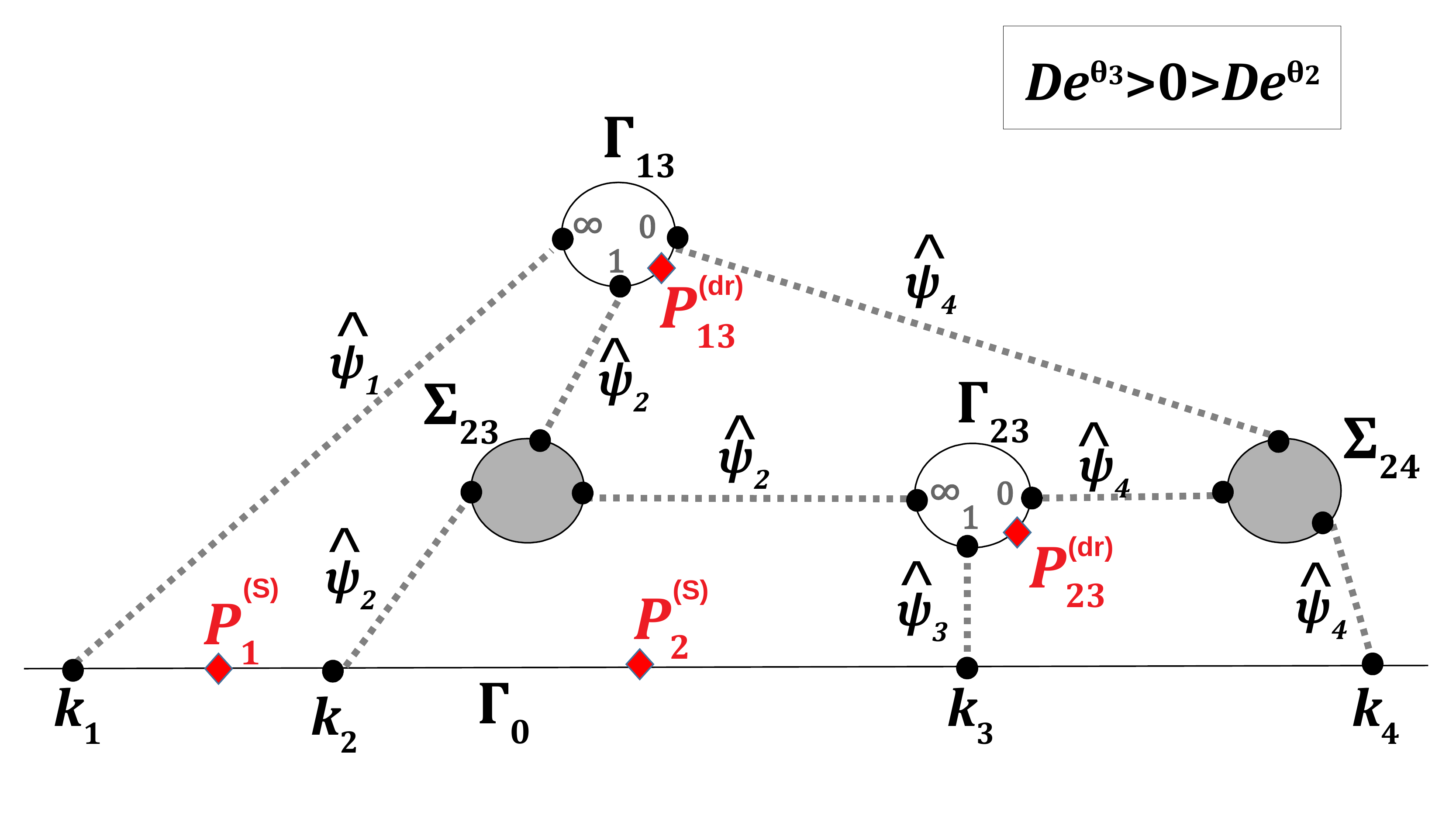}
\hfill
\includegraphics[width=0.3\textwidth]{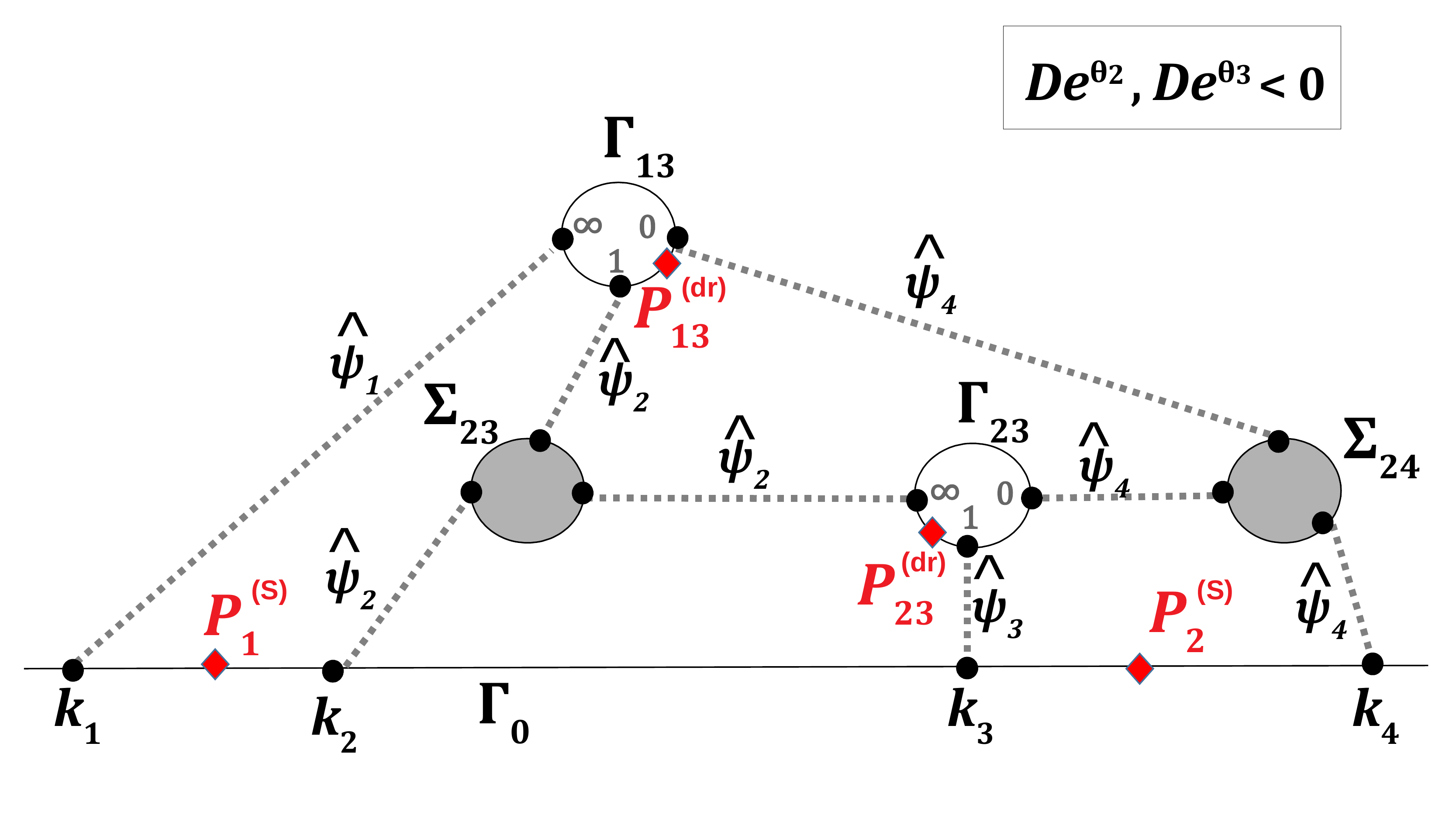}
\hfill
\includegraphics[width=0.3\textwidth]{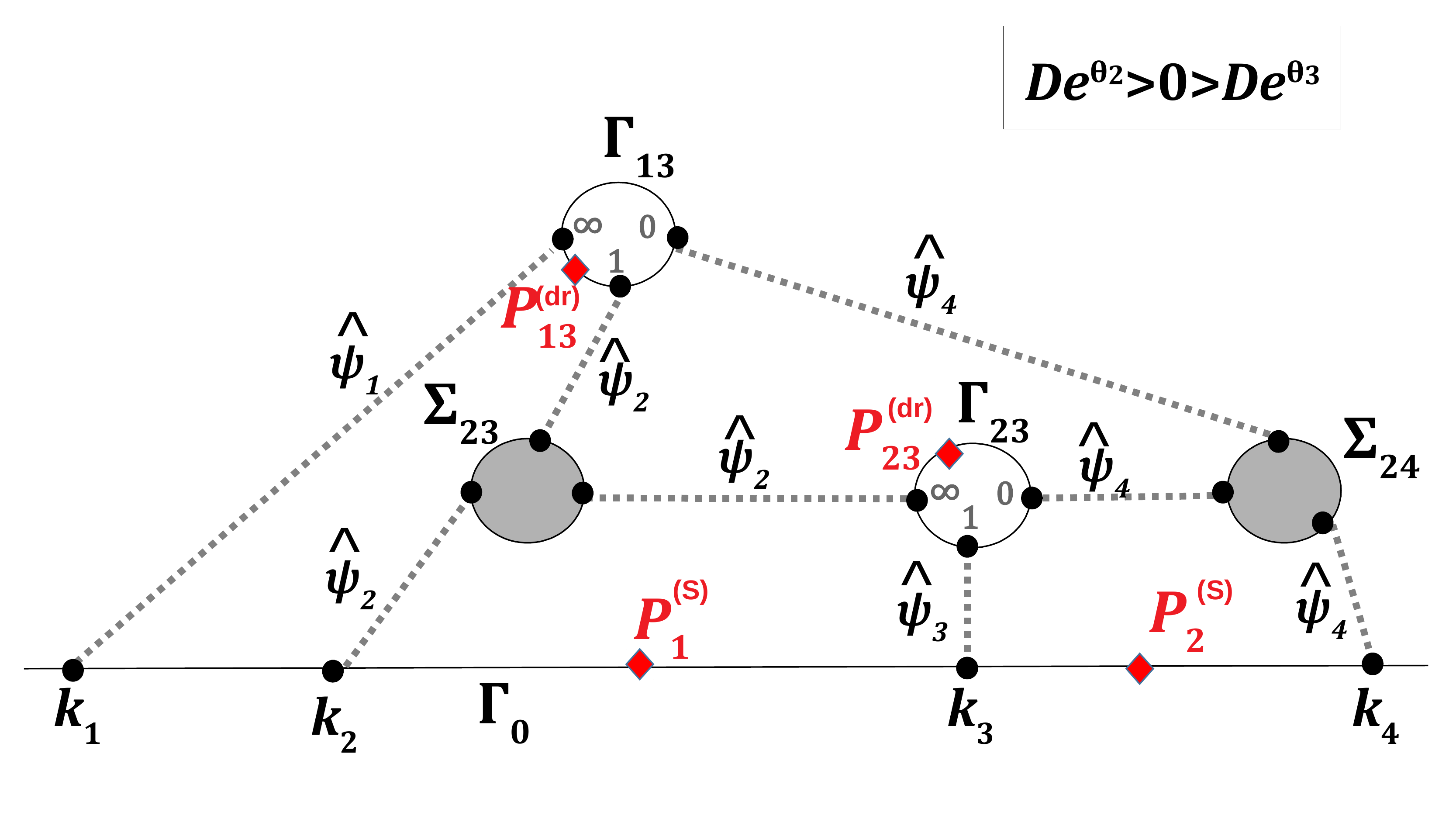}}
\caption{\small{\sl The pole divisor $\DKP$ is independent of the orientation. In the case of generic soliton data in $Gr^{\mbox{\tiny TP}}(2,4)$ there are only three possible configurations.}}
\label{fig:Gr24_pole}
\end{figure}

For the rest of the paper $\Gamma=\Gamma ({\mathcal N}_{T,{\mbox{\scriptsize red}}})$. 
By construction the KP divisor $\DKP$ consists of the degree $k=2$ Sato divisor $(P_1,P_2)$ defined in (\ref{eq:Satodiv}),
$\zeta(P_l)= \gamma_l$, $l=1,2$, and of $2$ simple poles $(P_{13},P_{23})$ respectively belonging to the intersection of 
$\Gamma_{13}$ or $\Gamma_{23}$ with the union of the finite ovals. In the local coordinates induced by the orientation of the Le--network \cite{A2,AG3}, we have
\[
\begin{array}{ll}
\zeta(P_1) +\zeta(P_2) = {\mathfrak w}_1 (\vec t_0), &\quad  \zeta(P_1)\zeta(P_2) = -{\mathfrak w}_2 (\vec t_0),\\
\zeta(P_{13}) = \frac{w_{14} {\mathfrak D} e^{\theta_4(\vec t_0)}}{{\mathfrak D} e^{\theta_3(\vec t_0)}+ (w_{14}+w_{24}) 
{\mathfrak D} e^{\theta_4(\vec t_0)}}, &\quad
\zeta(P_{23}) = \frac{w_{24} {\mathfrak D} e^{\theta_4(\vec t_0)}}{{\mathfrak D} e^{\theta_3(\vec t_0)}+ w_{24} 
{\mathfrak D} e^{\theta_4(\vec t_0)}}.
\end{array}
\]
For generic soliton data $[A]\in Gr^{\mbox{\tiny TP}}(2,4)$, the KP--II pole divisor configuration is one of the three shown in Figure \ref{fig:Gr24_pole} and it depends only on the signs of ${\mathfrak D} e^{\theta_2(\vec t_0)}$, ${\mathfrak D} e^{\theta_3(\vec t_0)}$, since ${\mathfrak D} e^{\theta_1(\vec t)}$, ${\mathfrak D} e^{\theta_4(\vec t)}>0$ and 
 ${\mathfrak D} f^{(1)}(\vec t)\equiv 0 \equiv {\mathfrak D} f^{(2)}(\vec t)$ for all $\vec t$ in agreement with \cite{Mal}, see also \cite{AG3}. For any given $[A]\in Gr^{\mbox{\tiny TP}}(2,4)$, in each finite oval, there is exactly one KP divisor point, where we use the counting rule established in \cite{AG1} for non--generic soliton data.

\section{The spectral curve for $Gr^{\mbox{\tiny TP}}(2,4)$ and its desingularization}

\begin{figure}[H]
  \centering
  {\includegraphics[width=0.44\textwidth]{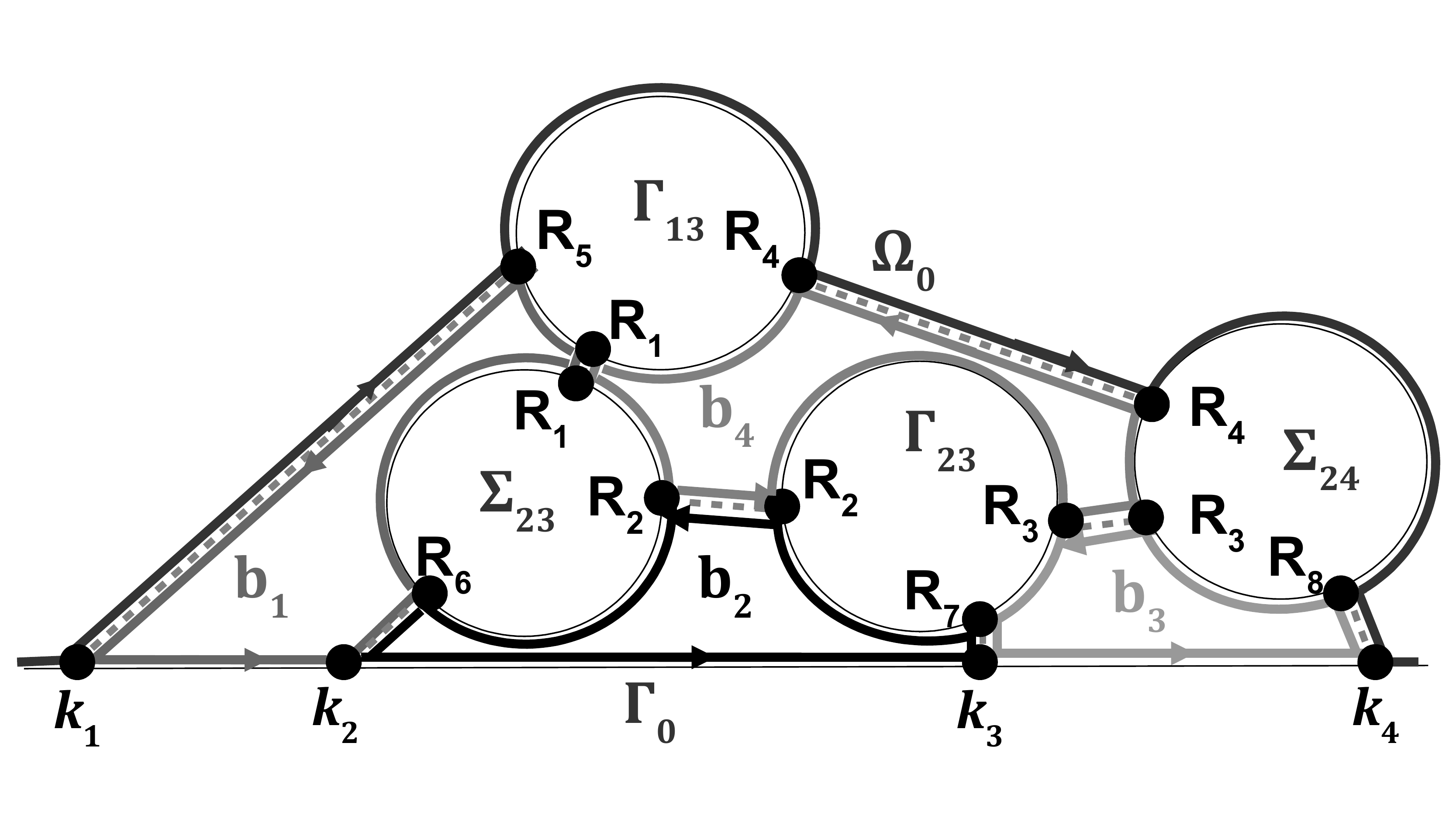}}
  \hspace{.6 truecm}
  {\includegraphics[,width=0.44\textwidth]{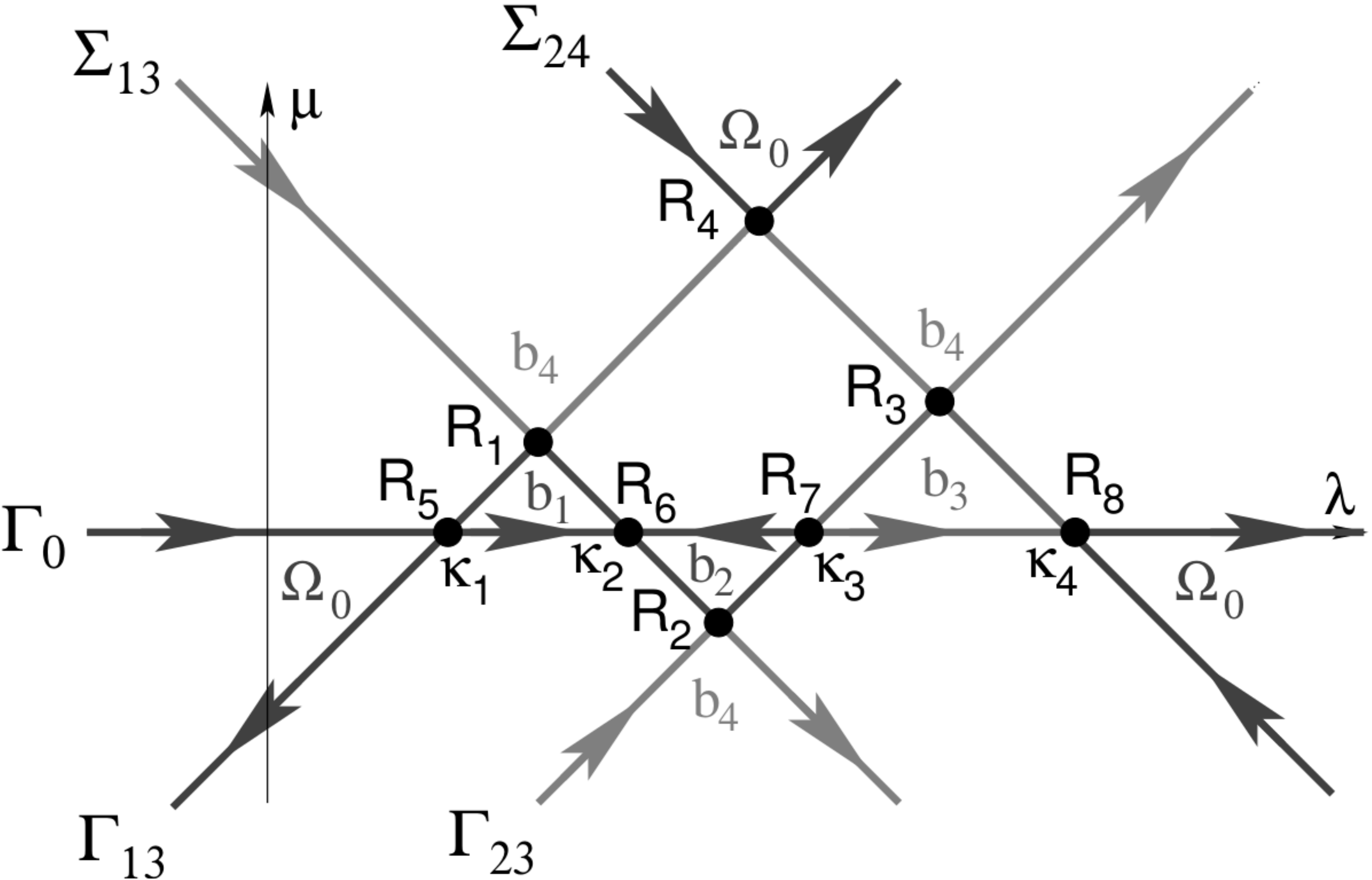}}
\caption{\small{\sl The topological scheme of spectral curve for soliton data $Gr^{\mbox{\tiny TP}}(2,4)$, $\Gamma({\mathcal N}_{T,{\mbox{\scriptsize red}}})$ (left) is the partial normalization the plane algebraic curve (right), which is a rational degeneration of the genus 4 $\mathtt M$--curve in (\ref{eq:curveGr24_pert}). The ovals in the nodal plane curve are labeled as in the real part of its partial normalization.}}\label{fig:gr24_top}        
\end{figure}

In this section we construct the rational curve associated to soliton data in $Gr^{\mbox{\tiny TP}} (2,4)$. The degenerate curve $\Gamma({\mathcal N}_{T,{\mbox{\scriptsize red}}})$ is the partial normalization of the nodal plane curve in (\ref{eq:curveGr24}), and it is the rational degeneration of the genus 4 $\mathtt M$--curve in (\ref{eq:curveGr24_pert}) when $\varepsilon\to 0$. 
Let us recall that generic curves of sufficiently high genus can not be represented as plane curves without self-intersections \cite{GrH}, \cite{ACG}, therefore the partial normalization is generically unavoidable. We plot both the topological model and the partial normalization for this example in Figure \ref{fig:gr24_top}. 

The degenerate curve $\Gamma({\mathcal N}_{T,{\mbox{\scriptsize red}}})$ is obtained gluing five copies of $\mathbb{CP}^1$: $\Gamma_0$, $\Gamma_{13}$, $\Gamma_{23}$, $\Sigma_{23}$ and $\Sigma_{24}$ and it may be represented as a plane curve given by the intersection of five lines. To simplify its representation, we impose that the line representing $\Gamma_0$ is one of the coordinate axis, that $P_0$ is the infinite point, and that the line representing $\Gamma_{13}$ is parallel to the one representing $\Gamma_{23}$, and orthogonal to those representing $\Sigma_{23}$ and $\Sigma_{24}$. Acting with the group of affine transformations on the $(\lambda,\mu)$-plane, we assume without loss of generality that the coordinate $\lambda$ coincides with $\zeta$ on $\Gamma_0$ and the 5 components are defined by the equations:
\begin{equation}\label{eq:lines}
\Gamma_0:  \mu=0, \ \ \Gamma_{13}: \mu-\lambda + \kappa_1 =0, \ \  \Gamma_{23}: \mu-\lambda +\kappa_3 =0, \ \ 
\Sigma_{13}: \mu + \lambda -\kappa_2=0, \ \ \Sigma_{24}: \mu + \lambda -\kappa_4=0.
\end{equation}

We assume that the singularity at infinity is completely resolved, therefore the lines $\Gamma_{13}$ and $\Gamma_{23}$ do not intersect at infinity, and, similarly,  $\Sigma_{23}$ and $\Sigma_{24}$ do not intersect at infinity. In agreement with the finite-gap approach, see Remark~\ref{rem:ovals}, we denote the finite ovals $b_j$, $j\in[4]$. Let us remark, that the oval $b_4$ is finite, because it does not pass through $P_0$.

Let us explain the relation between the coordinate $\lambda$ and the original coordinate $\zeta$ at each copy.

On $\Gamma_{13}$, we have 3 real ordered marked points, with $\zeta$--coordinates: $\zeta(R_4)=0<\zeta(R_1)=1<\zeta(R_5)=\infty$.
Comparing with (\ref{eq:lines}) we then easily conclude that
\[
\lambda = \frac{2\kappa_1 (\kappa_4-\kappa_2) \zeta +(\kappa_2-\kappa_1)(\kappa_1+\kappa_4)}{(\kappa_4-\kappa_2) \zeta +(\kappa_2-\kappa_1)}.
\]

On $\Sigma_{23}$, we have 3 real ordered marked points, and the following constraints: $\lambda(R_6)=\kappa_2$, $\lambda(R_1)=\frac{\kappa_1+\kappa_2}{2}$, $\mu(R_1)=\frac{\kappa_2-\kappa_1}{2}$.

Similarly on $\Sigma_{24}$, we have 3 real ordered marked points, and the following constraints: $\lambda(R_8)=\kappa_4$, $\lambda(R_4)=\frac{\kappa_1+\kappa_4}{2}$, $\mu(R_4)=\frac{\kappa_4-\kappa_1}{2}$.

Analogously, on $\Gamma_{23}$ in the initial $\zeta$ coordinates we have 3 real ordered marked points and $\zeta(R_3)=0<\zeta(R_7)=1<\zeta(R_2)=\infty$, therefore the fractional linear transformation to the $\lambda$ is:
\[
\lambda = \frac{(\kappa_2+\kappa_3) (\kappa_4-\kappa_3) \zeta +(\kappa_3-\kappa_2)(\kappa_3+\kappa_4)}{(\kappa_4-\kappa_3) \zeta +(\kappa_3-\kappa_2)}.
\]

Finally, $\Gamma({\mathcal N}_{T,{\mbox{\scriptsize red}}})$ is represented by the reducible plane curve $P_0(\lambda,\mu)=0$, with
\begin{equation}
\label{eq:curveGr24}
P_0(\lambda,\mu)=\mu\cdot\big(\mu-(\lambda-\kappa_1)\big)\cdot\big(\mu+(\lambda-\kappa_2)\big)\cdot\big(\mu-(\lambda-\kappa_3)\big)\cdot\big(\mu+(\lambda-\kappa_4)\big).
\end{equation}

To obtain an $\mathtt M$--curve after desingularization we have to check that each double point is opened in a correct way, which implies some inequality-type constraints on the perturbation. It is easy to check, that the following perturbation generates a smooth genus 4  $\mathtt M$--curve, $\Gamma_{\varepsilon}$:
\begin{equation}
\label{eq:curveGr24_pert}
\Gamma(\varepsilon) \; : \quad\quad P(\lambda, \mu)= P_0(\lambda,\mu) + \varepsilon (\beta^2 -\mu^2)=0, \quad\quad
0<\varepsilon \ll 1,
\end{equation}
where
\[
\beta = \frac{\kappa_4-\kappa_1}{4} +\frac{1}{4} \max \left\{  \kappa_2-\kappa_1, \kappa_3-\kappa_2, \kappa_4-\kappa_3  \right\},
\]
and $\Gamma({\mathcal N}_{T,{\mbox{\scriptsize red}}})$ is the rational degeneration of $\Gamma_{\varepsilon}$. 

\begin{figure}[H]
  \centering
  {\includegraphics[,width=0.6\textwidth]{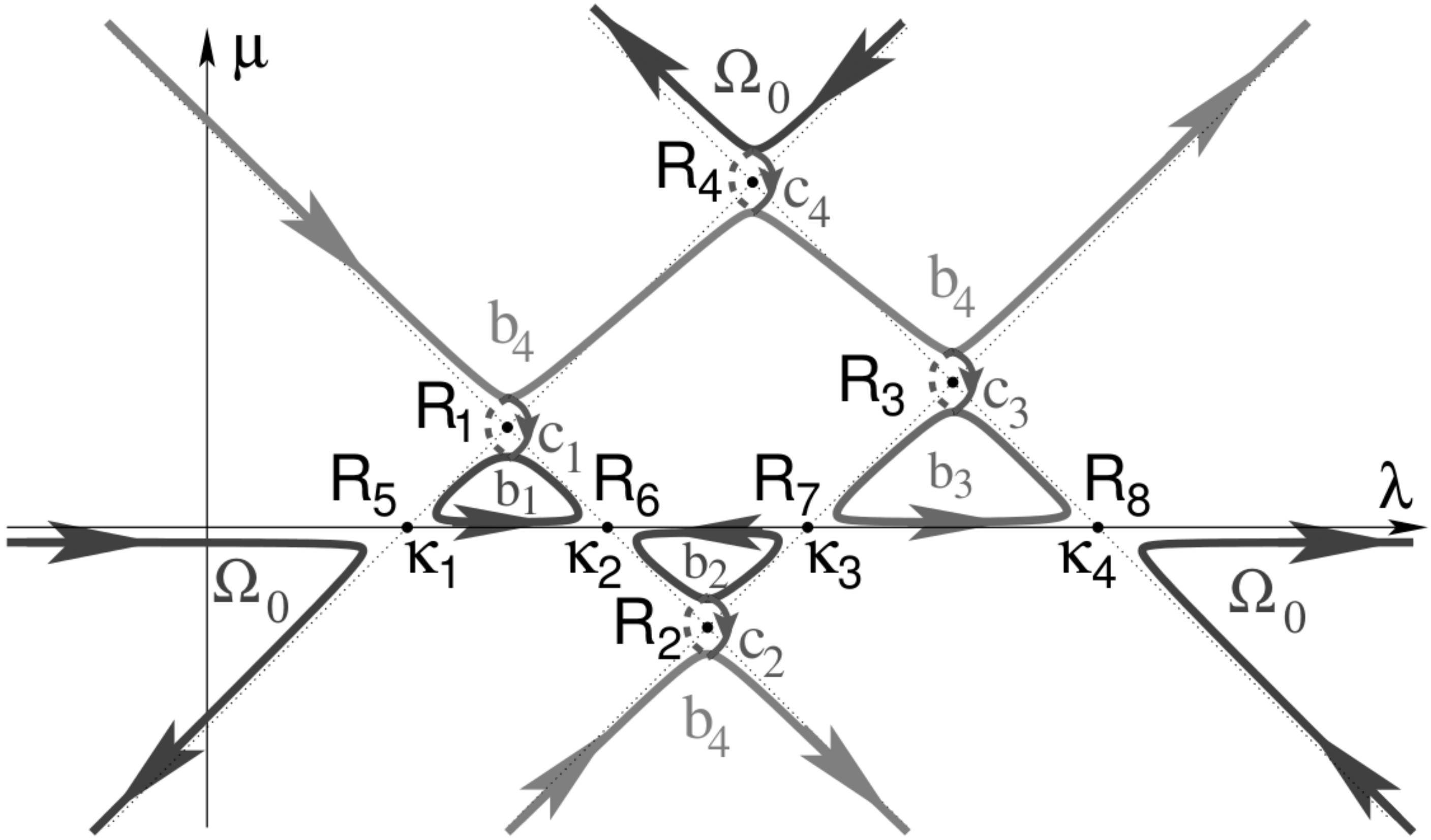}}
\caption{\small{\sl The desingularization  $\Gamma_{\varepsilon}$ of the spectral curve  $\Gamma({\mathcal N}_{T,{\mbox{\scriptsize red}}})$ for soliton data $Gr^{\mbox{\tiny TP}}(2,4)$. The $b$-cycles correspond to the finite ovals. On contours $c_j$ the solid lines correspond to $\Im\lambda>0$, the dashed lines correspond to $\Im\lambda<0$.}\label{fig:gr24_smooth}}        
\end{figure}

\section{The numerical simulation}

To illustrate the above construction we provide the results of numerical calculations for the following choice of parameters:
\[
\kappa_1 =- 1.5, \ \ \kappa_2 = -0.75, \ \ \kappa_3= 0.5, \ \ \kappa_4= 2.
\]
For the basis of cycles illustrated in Figure~\ref {fig:gr24_smooth} we have the following intersection numbers:
\begin{align}
c_1 \circ b_1 &=-1  & c_2 \circ b_1 &=0 & c_3 \circ b_1 &=0  & c_4 \circ b_1 &=0 \nonumber\\
c_1 \circ b_2 &=0   & c_2 \circ b_2 &=-1 & c_3 \circ b_2 &=0  & c_4 \circ b_2 &=0  \\
c_1 \circ b_3 &=0   & c_2 \circ b_3 &=0 & c_3 \circ b_3 &=-1  & c_4 \circ b_3 &=0 \nonumber \\
c_1 \circ b_4 &=-1  & c_2 \circ b_4 &=1 & c_3 \circ b_4 &=-1 & c_4 \circ b_4 &=1. \nonumber 
\end{align}
Therefore:
\[
a_1 = -c_1 - c_4, \ \ a_2 = -c_2 + c_4, \ \ a_3 = -c_3 - c_4, \ \ a_4 = c_4. 
\]
The basis of unnormalized holomorphic and meromorphic differentials are
\[
\sigma_1 = \frac{d\lambda}{P_\mu}, \ \ 
\sigma_2 = \frac{\lambda d\lambda}{P_\mu}, \ \
\sigma_3 = \frac{\mu d\lambda}{P_\mu}, \ \
\sigma_4 = \frac{(\lambda^2-\mu^2)d\lambda}{P_\mu}, \ \
\Sigma_1 = \frac{Q d\lambda}{P_\mu}, \ \
\Sigma_2 = \frac{\lambda Q d\lambda}{P_\mu}, \ \
\Sigma_3 = \frac{\lambda^2 Q d\lambda}{P_\mu},
\]
with
\[
Q(\lambda,\mu)=\big(\mu-(\lambda-\kappa_1)\big)\cdot\big(\mu+(\lambda-\kappa_2)\big)\cdot\big(\mu-(\lambda-\kappa_3)\big)\cdot\big(\mu+(\lambda-\kappa_4)\big) - \varepsilon \mu = \frac{P(\lambda,\mu)-\varepsilon \beta^2}{\mu}.
\]
Differentials $\sigma_j$, $j=1,\ldots,4$ are holomorphic. Differentials $\Sigma_j$,$j=1,\ldots,3$  are holomorphic outside the point $P_0$ where:
\[
\mu=\frac{-\varepsilon \beta^2}{(\lambda-\kappa_1)\cdot(\lambda-\kappa_2)\cdot(\lambda-\kappa_3)\cdot(\lambda-\kappa_4)}+O\left(\frac{1}{\lambda^{10}} \right).
\] 
On the curve near the point $P_0$ we have:
\[
Q(\lambda,\mu) = -\frac{\varepsilon\beta^2}{\mu}= (\lambda-\kappa_1)\cdot(\lambda-\kappa_2)\cdot(\lambda-\kappa_3)\cdot(\lambda-\kappa_4)+ \frac{\varepsilon\beta^2(\kappa_1+\kappa_3-\kappa_2-\kappa_4 )}{\lambda^{2}} +O\left(\frac{1}{\lambda^{3}} \right),
\]
\[
P_\mu =(\lambda-\kappa_1)\cdot(\lambda-\kappa_2)\cdot(\lambda-\kappa_3)\cdot(\lambda-\kappa_4) +\frac{2\varepsilon\beta^2(\kappa_1+\kappa_3-\kappa_2-\kappa_4 )}{\lambda^{2}} + O\left(\frac{1}{\lambda^{3}} \right),
\]
and
\begin{align}
\Sigma_1 &= \left(1 + \frac{\varepsilon\beta^2(\kappa_2+\kappa_4-\kappa_1-\kappa_3)}{\lambda^{6}} + O\left(\frac{1}{\lambda^{7}}\right) \right)d\lambda, \\
\Sigma_2 &= \left(\lambda +\frac{\varepsilon\beta^2(\kappa_2+\kappa_4-\kappa_1-\kappa_3)}{\lambda^{5}} + O\left(\frac{1}{\lambda^{6}}\right) \right)d\lambda, \\
\Sigma_3 &= \left(\lambda^2 + \frac{\varepsilon\beta^2(\kappa_2+\kappa_4-\kappa_1-\kappa_3)}{\lambda^{4}} +O\left(\frac{1}{\lambda^{5}}\right) \right)d\lambda.
\end{align}
Similarly one can easily write the expansions of the holomorphic differentials $\sigma_j$ near $P_0$ up to $O(1/\lambda^5)$ corrections. 

Using Gauss integrator with adaptive step, we calculate the integrals of the unnormalized holomorphic and meromorphic differentials over the cycles $b_j$, $c_j$, then calculate the canonical basis of holomorphic and meromorphic differentials, their periods and expansion coefficients near infinity up to $O(1/\lambda^5)$ corrections. Then we plot the graphs of $u(x,y,t)$ in the $(x,y)$ plane for $t$ fixed, using Its-Matveev formula. 

To guarantee numerical stability for almost degenerate curves, we calculate the parameters of the finite-gap solutions using quadruple precision.

To verify the consistency of our numerical calculation, we use the following tests:
\begin{enumerate}
\item We check the symmetry of the Riemann matrix. In the worst case $\epsilon=10^{-18}$ the matrix is symmetric up to $10^{-20}$ error.
\item We check that expansion coefficients of the Abel transform near infinity coincide with the $b$-periods of normalized meromorphic differential up to $2\pi i$ factor. This check is fulfilled with error less than $10^{-20}$.
\item We check the symmetry of the expansion matrix $\hat\omega_{jk}$. The non-symmetry is again less than $10^{-20}$. 
\item We write the theta-functional expansions for the derivatives of $u(\vec t)$, and we substitute them into KP-II equation. The error varies from $10^{-32}$ to $10^{-25}$ depending on the point in the $(x,y,t)$-space.
\end{enumerate}

\begin{figure}[H]
  \centering
  {\includegraphics[width=0.29\textwidth]{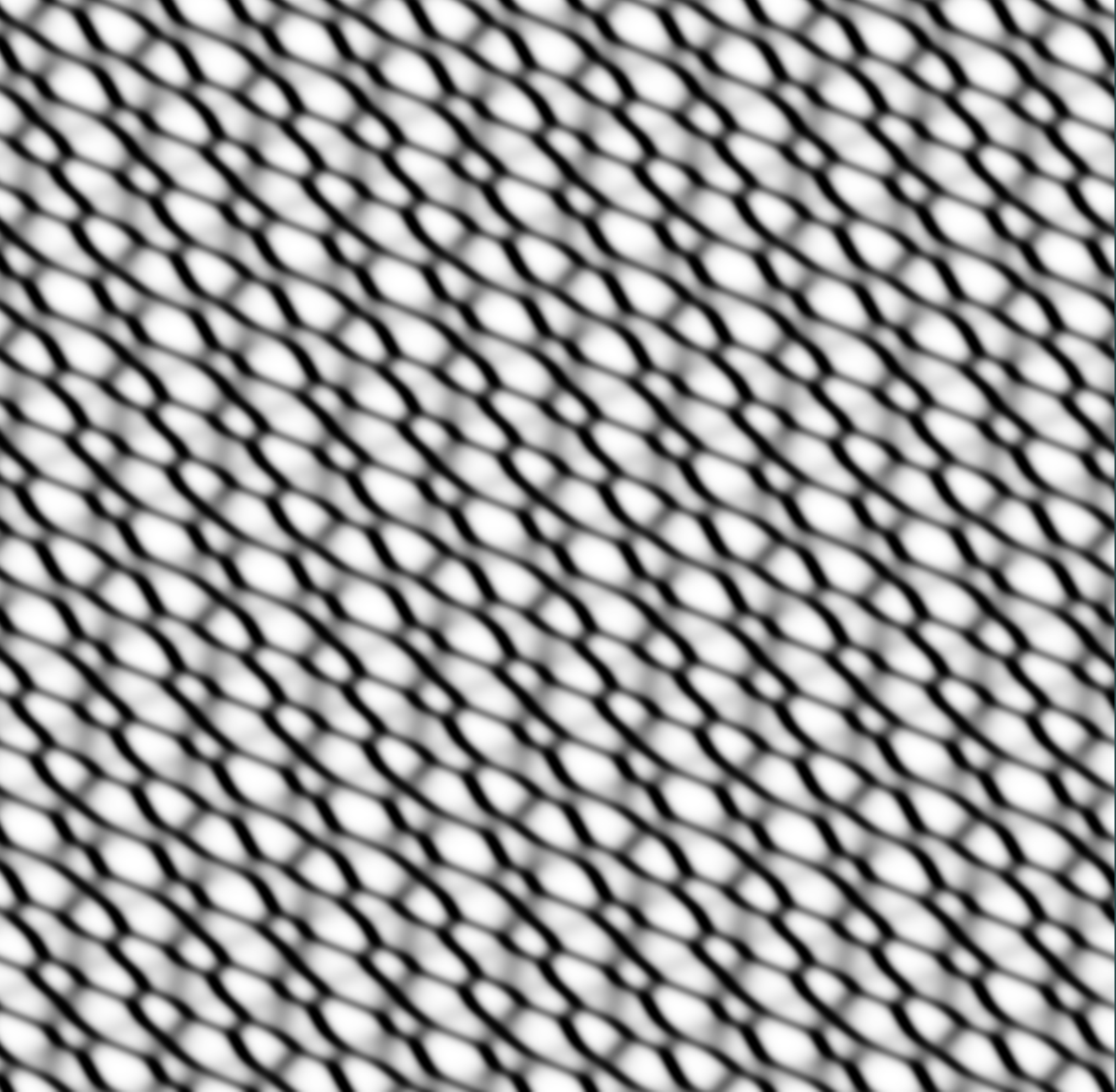}
  \hfill
  \includegraphics[width=0.29\textwidth]{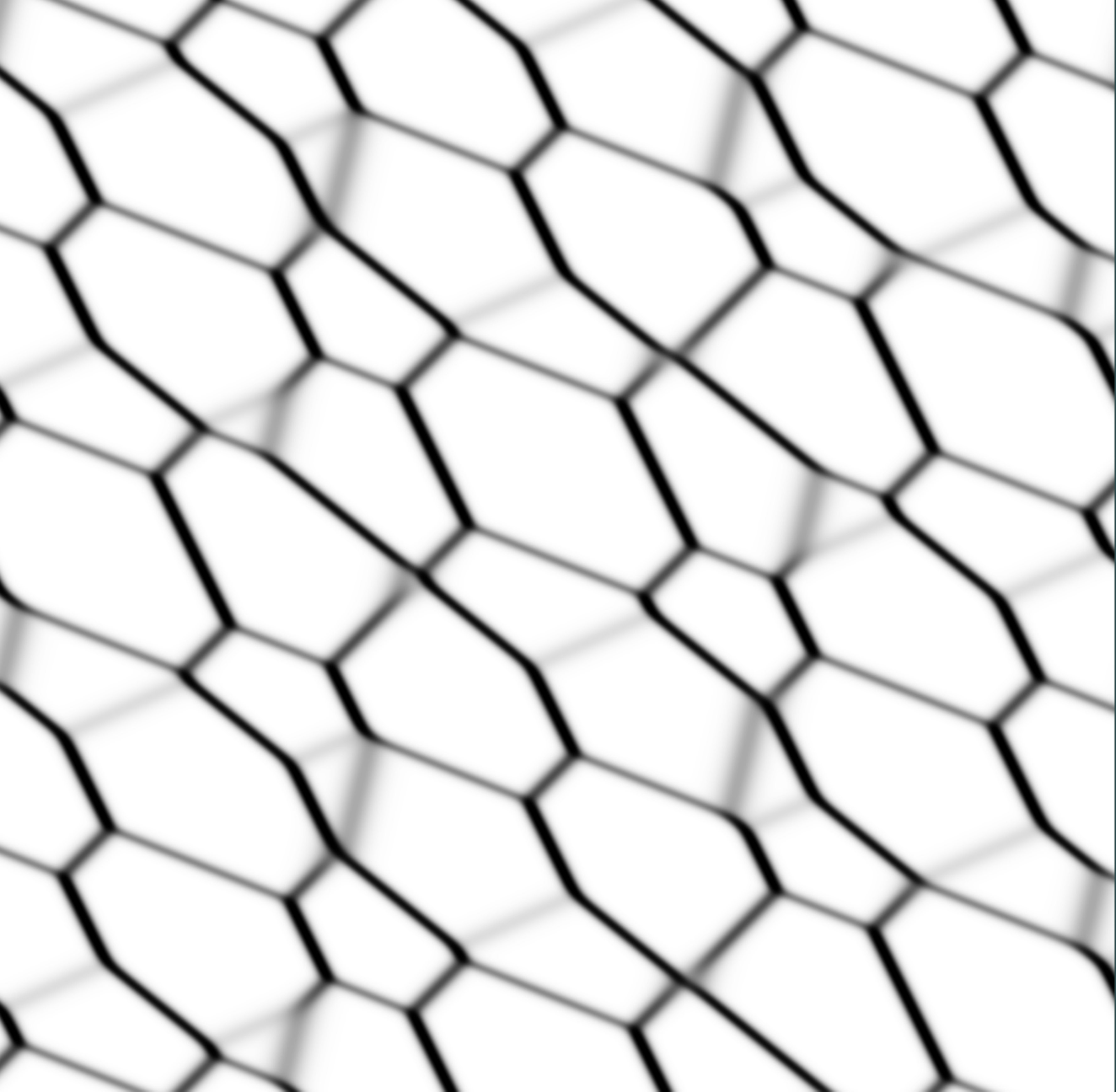}
  \hfill
  \includegraphics[width=0.29\textwidth]{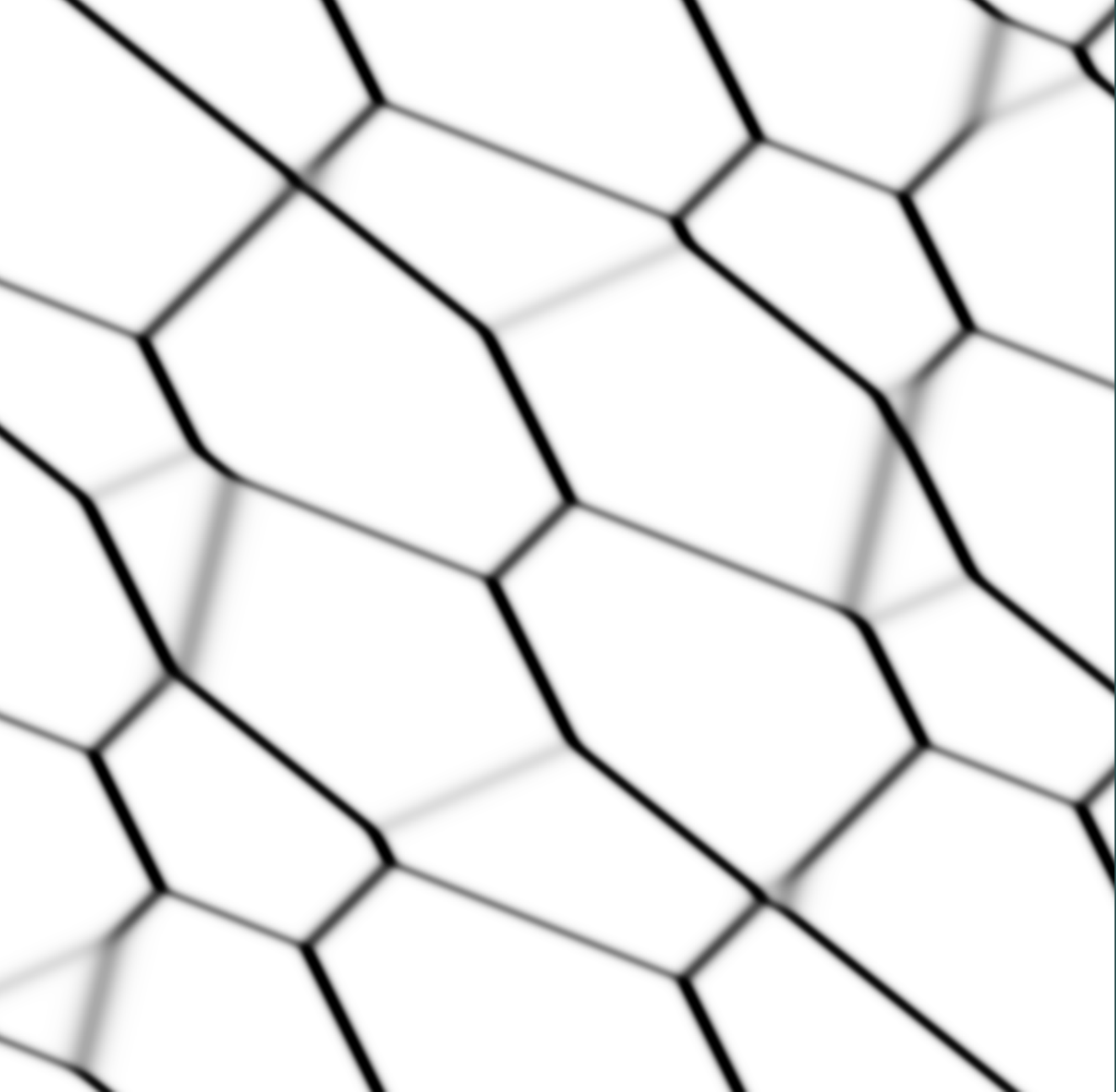}}
\caption{\small{\sl Level plots for the KP-II solutions for $\epsilon=10^{-2}$ [left], $\epsilon=10^{-10}$ [center] and 
$\epsilon=10^{-18}$ [right]. The horizontal axis is $-60\le x \le 60$, the vertical axis is $0\le y \le 120$, $t=0$. The white color corresponds to lowest values of $u$, the dark color corresponds to the highest values of $u$.}}\label{fig:level_plot}        
\end{figure}

\begin{figure}[H]
  \centering
  {\includegraphics[width=0.29\textwidth]{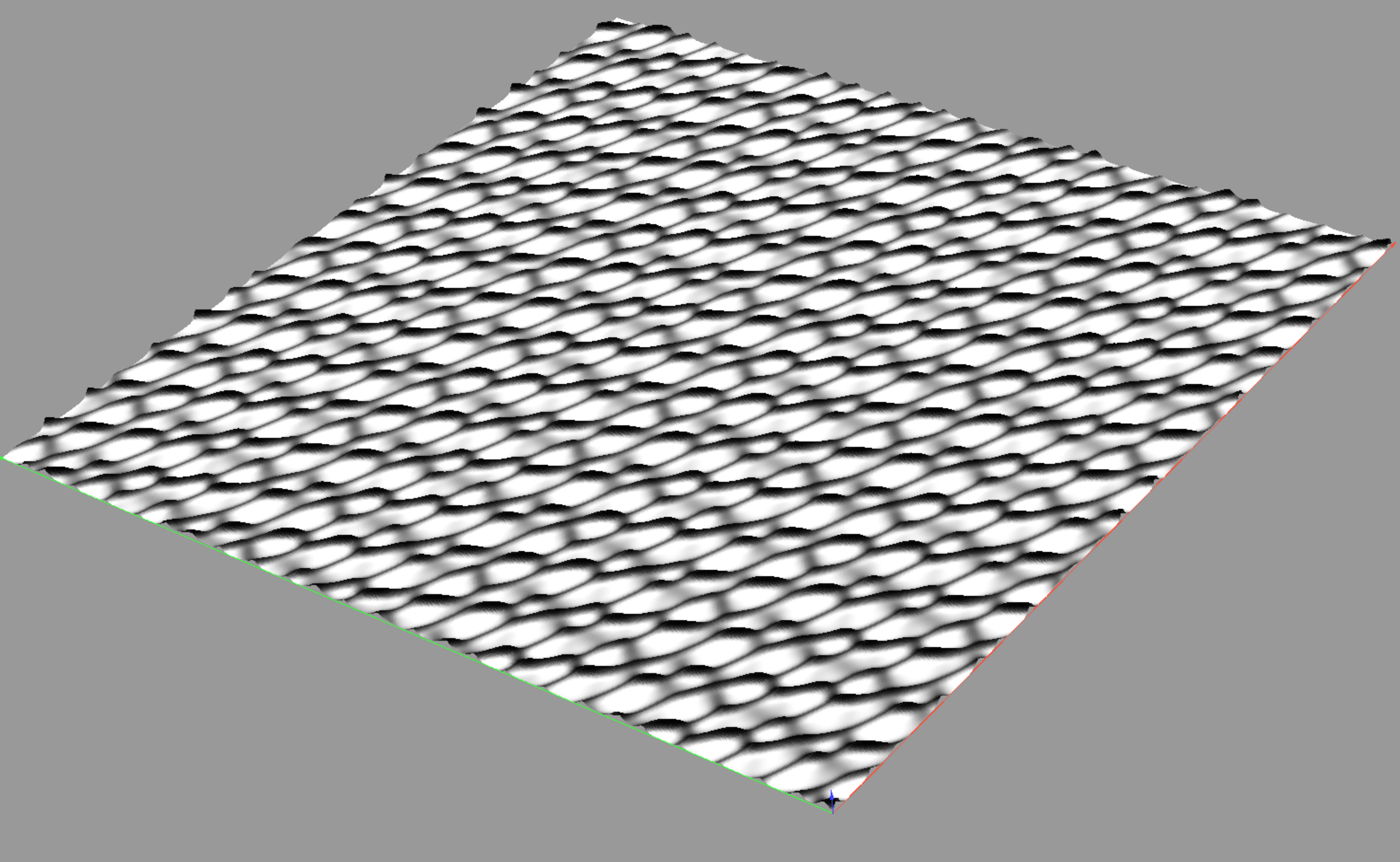}
  \hfill
  \includegraphics[width=0.29\textwidth]{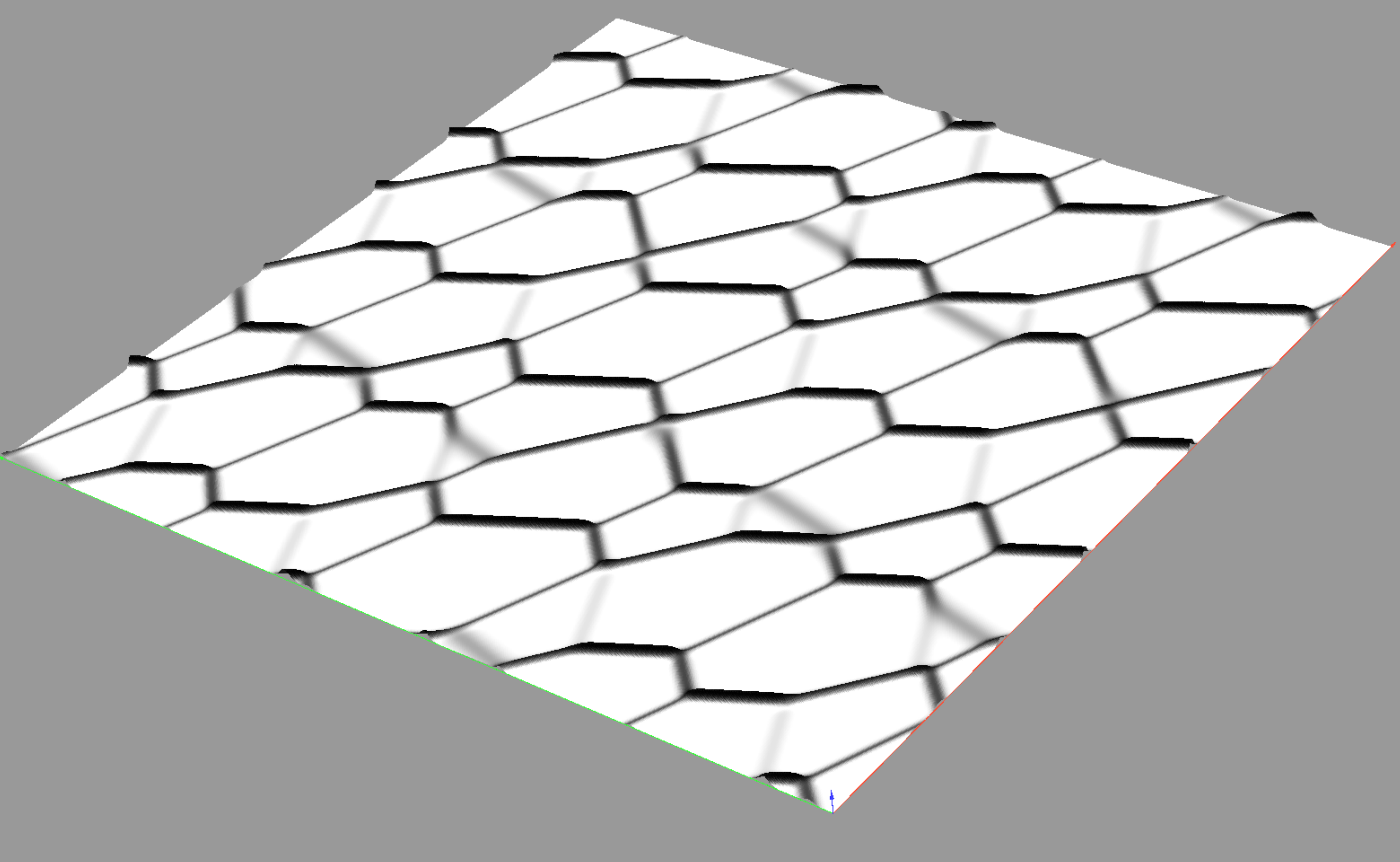}
  \hfill
  \includegraphics[width=0.29\textwidth]{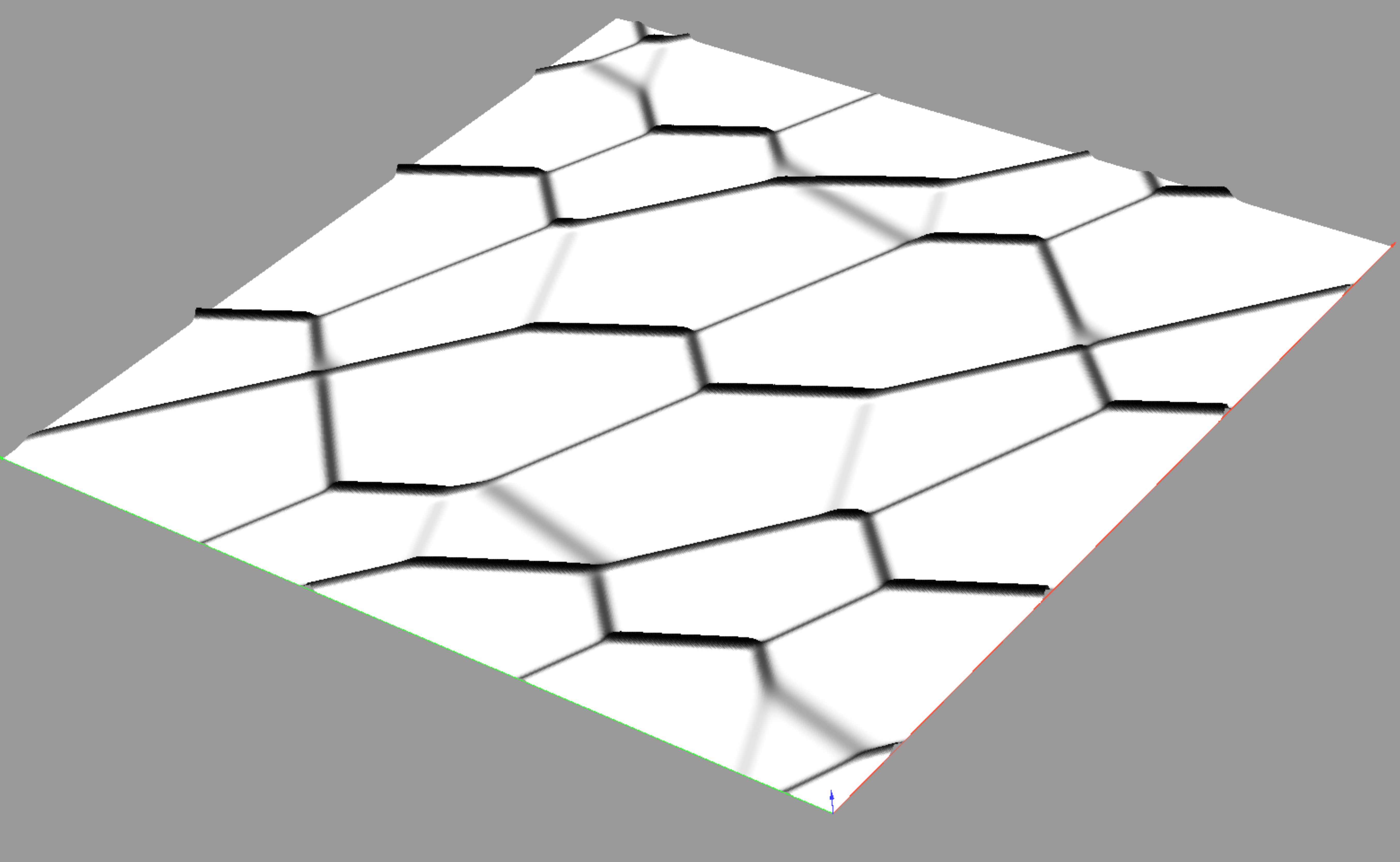}}
\caption{\small{\sl 3D-plots for the KP-II solutions.  The parameters and colors are the same as in Figure~\ref{fig:level_plot}}}\label{fig:level_3D}        
\end{figure}

In figures \ref{fig:level_plot} and \ref{fig:level_3D} we present the plots obtained by numerical simulations for different values of $\epsilon$.


\bibliographystyle{alpha}

\end{document}